\documentstyle[12pt,aaspp4]{article}
\begin{document}
\title{Ionization Sources and Physical Conditions in the Diffuse Ionized Gas Halos of Four Edge-On Galaxies}
\author{Joseph A. Collins and Richard J. Rand\altaffilmark{1}}
\affil{University of New Mexico, Dept. of Physics and Astronomy, 800 Yale Blvd.
 NE, Albuquerque, NM 87131}
\altaffiltext{1}{Visiting Astronomer, Kitt Peak National Observatory, National Optical Astronomy Observatories, Tucson, AZ}

\begin{abstract}
Deep long-slit spectra of the diffuse ionized gas halos of the edge-on spiral galaxies NGC 4302 and UGC 10288 are presented.  Emission lines are detected up to about $z=2$ kpc in NGC 4302, and to nearly $z=3$ kpc on the north side of UGC 10288.  For both galaxies, the line ratios [\ion{N}{2}]/H$\alpha$ and [\ion{S}{2}]/H$\alpha$ increase with $z$ in accordance with dilute photo-ionization models.  Runs of [\ion{S}{2}]/[\ion{N}{2}], and for UGC 10288, the run of [\ion{O}{3}]/H$\alpha$, however, are not explained by the models.  Scale height determinations of their DIG halos are generally lower than those of galaxies with more prominent extraplanar DIG features. 

These data, along with previously presented data for NGC 5775 and NGC 891, are used to address the issue of how DIG halos are energized.  Composite photo-ionization/shock models are generally better at explaining runs of line ratios in these galaxies than photoionization models alone.  Models of line ratios in NGC 5775 require a greater contribution from shocks for filamentary regions than for non-filamentary regions to explain the run of [\ion{O}{3}]/H$\alpha$.  In either case, the [\ion{S}{2}]/[\ion{N}{2}] ratio is not well fit by the models.  Composite models for UGC 10288 are successful at reproducing the run of [\ion{S}{2}]/[\ion{N}{2}] for all but the the highest values of [\ion{N}{2}]/H$\alpha$;  however, the run of [\ion{O}{3}]/H$\alpha$ vs. [\ion{N}{2}]/H$\alpha$ does not show any discernible trend, making it difficult to determine whether or not shocks contribute to the layer's maintenance.  

We also examine whether the data can be explained simply by an increase in temperature with $z$ in a pure photo-ionization model without a secondary source of ionization. Runs of [\ion{S}{2}]/H$\alpha$, [\ion{N}{2}]/H$\alpha$, and [\ion{S}{2}]/[\ion{N}{2}] in each of the four galaxies are consistent with such an increase.  However, the runs of [\ion{O}{3}]/H$\alpha$ vs. $z$ in NGC 5775 and UGC 10288 require unusually high ionization fractions of O$^{++}$ that can not be explained without invoking a secondary ionization source or at the very least a much higher temperature for the [\ion{O}{3}]-emitting component than for the [\ion{S}{2}]- and [\ion{N}{2}]-emitting component.  An increase in temperature with $z$ is generally more successful at explaining the [\ion{O}{3}]/H$\alpha$ run in NGC 891, with the ionization fraction of O$^{++}$ remaining relatively low and nearly constant with $z$. 

\end{abstract}


\section{Introduction}

A vertically extended layer known as the Reynolds Layer or the Warm Ionized Medium (WIM) of the Milky Way contains the majority of the ionized gas in the interstellar medium.  The layer fills about 20\% of the ISM volume with a local midplane density of $\sim$0.1 cm$^{-3}$ and a scale height of $\sim$1 kpc \markcite{}(Reynolds 1993).  Energetic arguments favor photo-ionization by massive stars in the disk for the layer's maintenance, as roughly 15\% their ionizing output is sufficient to maintain the layer in the solar neighborhood \markcite{}(Reynolds 1993).  The Wisconsin H$\alpha$ Mapper (WHAM; \markcite{}Reynolds et. al. 1998b) has been instrumental in recent years in characterizing the WIM.  In external galaxies, where the layer is more commonly referred to as Diffuse Ionized Gas (DIG), analogous information has been obtained through the use of narrow-band emission-line imaging, long-slit spectroscopy, and Fabry-Perot observations [e.g. Rand, Kulkarni, \& Hester 1990 (hereafter RKH); Rand 1996, 1997, 1998; Golla, Dettmar, \& Domg\"{o}rgen 1996; Ferguson, Wyse, \& Gallagher 1996; Wang, Heckman, \& Lehnert 1997; Hoopes, Walterbos, \& Rand 1999; Collins et. al. 2000 (hereafter CRDW)].  Estimates of the diffuse fraction in DIG layers put the ionization requirement closer to 40\% of the massive stars' ionizing output (e.g. Hoopes et. al. 1996; Ferguson et. al. 1996).  The advantage in observing external edge-on galaxies is the ability to obtain an overview of the entire halo, so that brightness, spatial distribution, and emission line ratios can be better characterized as they vary with $z$. 

One of the most important issues regarding these layers is their ionization.  In the WIM, as well as in external DIG layers, [\ion{S}{2}] $\lambda6716/$H$\alpha$ and [\ion{N}{2}] $\lambda6583/$H$\alpha$ are enhanced relative to [\ion{H}{2}] regions [e.g. Haffner, Reynolds, \& Tufte 1999 (hereafter HRT); Ferguson, Wyse, \& Gallagher 1996; Rand 1997].  Such a scenario is predicted by photo-ionization models in which the dilution of the radiation field, measured by the ionization parameter, $U$, increases as photons propagate away from \ion{H}{2}\ regions [Domg\"{o}rgen \& Mathis 1994; Sokolowski 1994 (hereafter S94); Bland-Hawthorn et. al. 1997].  Species such as S which are predominantly doubly ionized in \ion{H}{2}\ regions are able to recombine to a singly ionized state as the field becomes increasingly dilute.  Such an effect should not be as significant for N or O, which have considerably higher second ionization potentials and thus may not be fully doubly ionized in \ion{H}{2}\ regions. 

The photo-ionization scenario is not without problems, however.  Firstly, in an increasingly dilute field, photoionization models predict that the ratio [\ion{S}{2}]/[\ion{N}{2}] should increase with $z$ due to differing second ionization potentials of S and N.  However, [\ion{S}{2}]/[\ion{N}{2}] vs. $z$ has been found to remain nearly constant in the WIM \markcite{}(HRT) and NGC 891 \markcite{}(Rand 1998), and in the case of NGC 4631, the variations with $z$ are not as pronounced as those predicted by the models \markcite{}(Golla et al. 1996).  In addition, the ratio [\ion{O}{3}] $\lambda5007/$H$\alpha$ should decrease with $z$ in a photo-ionization scenario as the field becomes increasingly dilute.  The opposite trend, however, is seen in NGC 891 (Rand 1998), where [\ion{O}{3}]/H$\alpha$ increases significantly from the midplane to $z=2$ kpc, as well as in NGC 5775 (Rand 2000; T\"{u}llmann et. al. 2000).    [\ion{O}{3}]/H$\alpha$ enhancements in DIG relative to \ion{H}{2}\ regions are also seen in a number of more face-on galaxies, including M51 \markcite{}(Wang et. al. 1997).  Finally, photoionization models have trouble explaining values of [\ion{N}{2}]/H$\alpha$ and [\ion{S}{2}]/H$\alpha$ that are often $\gtrsim1$ (Wang et. al. 1997; Rand 1998; HRT).  Models of \markcite{}S94, who took observed line ratios in NGC 891 into consideration, can only explain such ratios by assuming an upper mass cutoff for the stellar IMF of $120 M_{\sun}$, hardening of the radiation field as it propagates, and a depletion of important gas-phase coolants onto dust grains.  Observations of the weak \ion{He}{1}\ $\lambda5876$ line, however, in the Reynolds Layer (Reynolds \& Tufte 1995) and in NGC 891 (Rand 1997) each imply a significantly softer spectrum than indicated by the forbidden lines.

One possibility that has been forwarded to explain some of this line ratio behavior is to assume a secondary ionizing source that contributes a small but increasing fraction of the H$\alpha$ emission with $z$.  Such secondary sources include shocks \markcite{}(Shull \& McKee 1979) or turbulent mixing layers (TMLs; \markcite{}Slavin, Shull, \& Begelman 1993), each of which may produce large amounts of [\ion{O}{3}] emission.  Other, less well-studied ionization sources may be relevant in halos, such as ionization from soft X-rays or cosmic rays.

Another approach that has been taken is to assume, instead of the ionization parameter or state changing with $z$, that the line ratio behavior can be explained by the gas temperature increasing with $z$.  This requires a source of additional heating, such as photo-electric heating from dust grains (Reynolds \& Cox 1992) or dissipation of interstellar turbulence (Minter \& Balser 1997), that becomes more significant at lower gas densities, but may not require any additional source of ionization.  A temperature increase with $z$ is successful in explaining enhancements in [\ion{N}{2}]/H$\alpha$ and [\ion{S}{2}]/H$\alpha$ with $z$, as well the constancy in [\ion{S}{2}]/[\ion{N}{2}] in the Reynolds Layer up to $z=1.75$ kpc (HRT; Reynolds, Haffner, \& Tufte 1999).  Such a rise in temperature could also explain an increase in [\ion{O}{3}]/H$\alpha$, though data on the run of [\ion{O}{3}]/H$\alpha$ vs. $z$ in the WIM are currently lacking.  

The goal of this research is to attempt to better understand how DIG halos are energized.  We attempt to determine whether measured runs of line ratios versus $z$ can be explained by: (\em{a}\em) photo-ionization with a contribution from shock ionization that increases with height above the midplane, and/or (\em{b}\em) only photo-ionization, but with an increase in gas temperature with $z$.  We examine our data in the context of each of these scenarios, realizing that elements of both could be important for understanding the ionization of DIG halos in general.   To this end, we have obtained data for two galaxies: NGC 4302 and UGC 10288.  We also employ data for NGC 5775, already presented by Rand (2000).  For the purpose of assessing the rising temperature scenario, we use data for NGC 891, which has been compared with composite shock/photo-ionization models previously \markcite{}(Rand 1998).  The observations of these galaxies are discussed in \S\ 2.  Line ratio data are presented in \S\ 3 along with comparisons to composite shock/photo-ionization and rising temperature models. We briefly summarize our results in \S\ 4. 

\section{Observations}
Observational details for the NGC 891 (Rand 1998) and NGC 5775 (Rand 2000) data have been discussed in previous papers, though some modifications have been made which will be discussed in this section.  Spectra for UGC 10288 and NGC 4302 were obtained at the KPNO 4-m telescope on 1999 June 10-13 and 1998 February 23-26, respectively.  Slits for all four galaxies run perpendicular to the plane. The slit position for NGC 4302, covering the most prominent region of extraplanar DIG, is shown in Figure 1, overlaid on the H$\alpha$ image.  The H$\alpha$ image of UGC 10288, showing little extraplanar H$\alpha$ emission, is shown in Figure 2 with slit position overlays.  The slit positions for UGC 10288 were chosen to be roughly 30\arcsec\ east and west of the galaxy's center.  Due to the presence of a foreground star on the axis perpendicular the galactic disk 30\arcsec\ west of the radio continuum center, slit 2 was placed at a position 35\arcsec\ west of center.  A summary of the observations is given in Table 1.

\begin{figure}
\figurenum{1}
\plotone{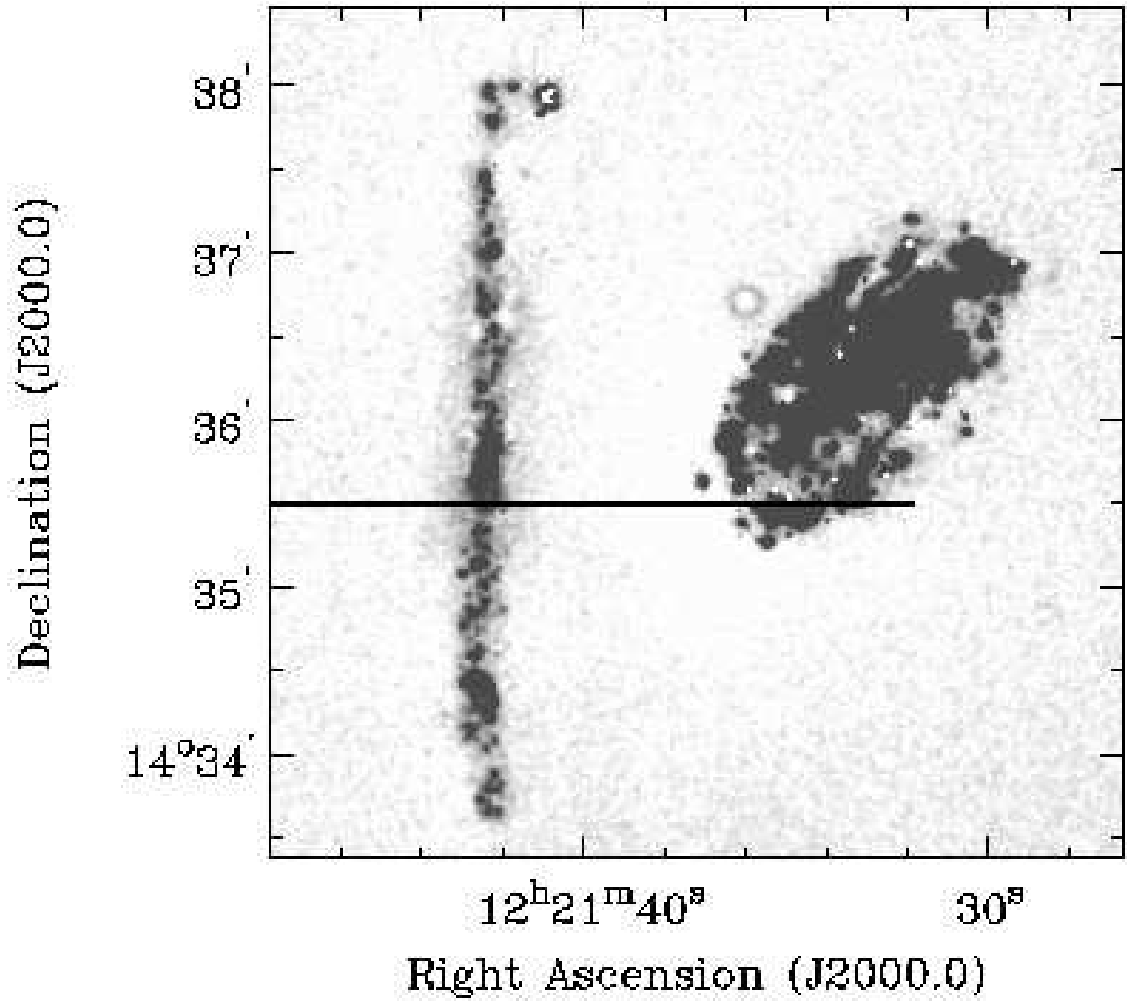}
\caption{The slit position for the observations of NGC 4302 overlayed on the H$\alpha$ image of Rand (1996). The image does not show the entire 5$\arcmin$ slit length.}
\end{figure} 

\begin{figure}
\figurenum{2}
\plotone{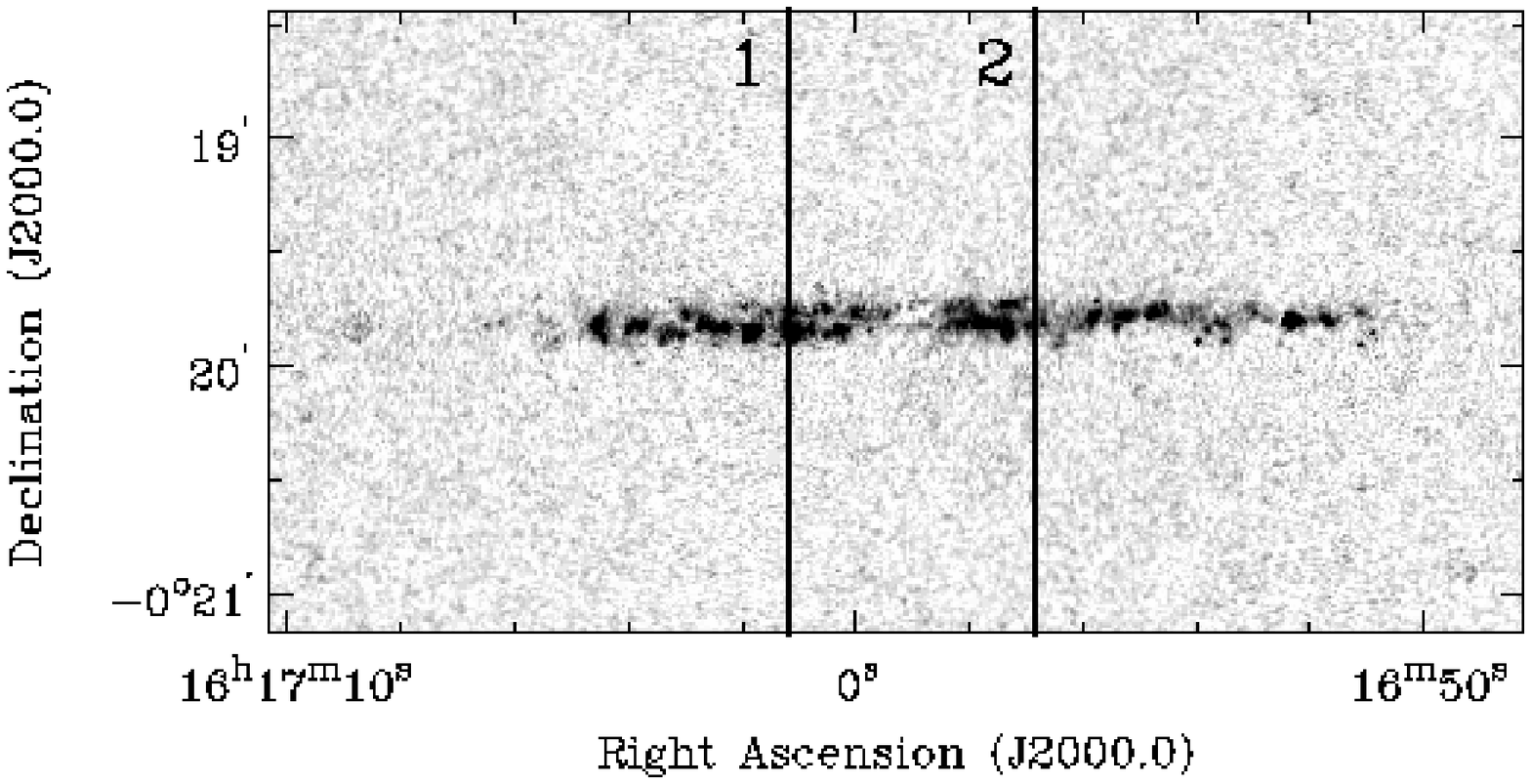}
\caption{The slit positions for the observations of UGC 10288 overlayed on the H$\alpha$ image of Rand (1996).  The image does not show the entire 5$\arcmin$ slit length.}
\end{figure}

\placetable{t1}

The observations of NGC 4302 utilized the KPC-24 grating in conjunction with the T2KB 2048x2048 CCD, providing a dispersion of 0.51\AA\ per pixel, a resolution of 1.3\AA\, and a useful coverage of about 800\AA.  The slit length is 5', the slit width was set at 3'', and the spatial scale is 0.69'' per pixel. For red line observations, the grating was tilted to give an approximate central wavelength of 6600\AA, allowing H$\alpha$, [\ion{N}{2}] $\lambda\lambda$6548, 6583, [\ion{S}{2}] $\lambda\lambda$6716, 6731, and [\ion{O}{1}] $\lambda$6300 to be observed.  Due to poor weather during the 1998 observing run, spectra containing [\ion{O}{3}] $\lambda$5007 could not be obtained.

For observations of UGC 10288, the KPC-007 grating was used with the T2KB 2048x2048 CCD, providing a dispersion of 1.42\AA\ per pixel, a resolution of 3.5\AA\, and a useful coverage of about 2100\AA\, encompassing the spectral lines H$\alpha$, [\ion{N}{2}] $\lambda\lambda$6548, 6583, [\ion{N}{2}] $\lambda$5755, [\ion{S}{2}] $\lambda\lambda$6716, 6731, \ion{He}{1}\ $\lambda$5876, [\ion{O}{1}] $\lambda$6300, [\ion{O}{3}] $\lambda$5007, and H$\beta$.  The slit length is 5', the slit width was set at 2'', and the spatial scale is 0.69'' per pixel.  

Data reductions were performed with the IRAF\footnotemark \footnotetext{IRAF is distributed by the National Optical Astronomy Observatories, which is operated by the Association for Research in Astronomy, Inc., under cooperative agreement with the National Science Foundation.} package.  Projector flats were used to remove small-scale variations in response, while sky flats were used to determine the slit illumination correction.  Arc lamp exposures were used to calibrate the wavelength scale as a function of location along the slit, and the spectral response function was determined with standard stars.  Regions of the spectra free of emission from the galaxies were used to form sky-subtracted spectra.  In order to increase sensitivity for the determination of line parameters, blocks of 5 pixels were averaged in the spatial direction.  Since Gaussians appear to accurately represent the line profiles, Gaussian fits were used to determine line properties.  Uncertainties in line parameters for NGC 4302 were determined with the IRAF task \it{splot}\rm.  The use of this method for NGC 5775 and UGC 10288 yielded error bars that were far greater than uncertainties suggested by the data.  In this case error bars were estimated by determining flux variations along the slit for sky lines of a range of intensities, from which a relation between these variances and sky line intensity could be established.  This relation was then used to determine flux uncertainties for the various emission lines.  Uncertainties for H$\alpha$ line strengths in NGC 5775 also take into account the coincidence of the H$\alpha$ line with a blended pair of sky lines.  This factor was not considered in the presentation of NGC 5775 line data in Rand (2000).  

In order to compare [\ion{O}{3}] $\lambda$5007 emission to that of H$\alpha$ it was necessary to make a reddening correction to the [\ion{O}{3}] line in NGC 5775, UGC 10288, and NGC 891.     This correction was not performed for the presentation of the [\ion{O}{3}]/H$\alpha$ ratio in Rand (2000) for NGC 5775.  In all three galaxies, H$\beta$ could be measured for all points where [OIII]$\lambda5007$ could be measured, thus allowing extinction corrections to be made for the full range in $z$. The correction was determined by assuming a uniform foreground dust screen and comparing measured values of the H$\beta$/H$\alpha$ ratio to the value calculated by Osterbrock (1989) for a 10$^{4}$ K gas. The assumption of a 10$^{4}$ K gas is not contradictory to a scenario including a rising temperature with $z$ in that the H$\beta$/H$\alpha$ ratio is not nearly as temperature sensitive as the ratios of forbidden lines to Balmer lines:  from 5,000 to 10,000 K, H$\beta$/H$\alpha$ increases by $\approx5\%$, whereas the ratios of [\ion{N}{2}], [\ion{S}{2}], and [\ion{O}{3}] to H$\alpha$ increase by an order of magnitude.   In NGC 5775, the correction to the [OIII] flux decreases from a factor of 2.9 in the midplane to no measurable extinction by $z=3$ kpc.  The correction is more variable for UGC 10288, with a maximum correction of 3.6 at $z=500$ pc on the south side.  In NGC 891, the correction decreases from 2.2 in the midplane to 1.5 at $z=1.6$ kpc.  In both NGC 5775 and NGC 891, the H$\beta$ line is superposed on a weak stellar absorption line close to the midplane.  However, the absorption line is much broader than the emission line and thus can be accounted for by the background fitting process in the line fitting routine. 

\section{Results and Discussion}
\subsection{Line Data and Line Ratios}
The H$\alpha$, [\ion{N}{2}] $\lambda$6583, and [\ion{S}{2}] $\lambda$6716 lines are detected up to nearly $z=2$ kpc on either side of the plane of NGC 4302.  The [\ion{S}{2}] $\lambda$6716 line was not detected at $z=1.6$ kpc on the west side, though the fainter line at 6731\AA\ was detected, due to confusion with noise in that part of the spectra.  The flux of the 6716\AA\ line at that part of the spectra was inferred from that of the 6731\AA\ line assuming the low-density limit of [\ion{S}{2}] $\lambda6716/\lambda6731=1.4$.  The vertical runs of [\ion{N}{2}] $\lambda6583/$H$\alpha$, [\ion{S}{2}] $\lambda6716/$H$\alpha$, and [\ion{S}{2}] $\lambda6716/$[\ion{N}{2}] $\lambda6583$  are shown in Figure 3.  The data are averaged over 5 spatial pixels.  The [\ion{N}{2}]/H$\alpha$ and [\ion{S}{2}]/H$\alpha$ each show a similar rise with $z$, with the [\ion{N}{2}]/H$\alpha$ rising from about 0.4 in the midplane to 1.0 at $z=1.2$ kpc on the east side, and to nearly 1.4 at $z=1.5$ kpc on the west side.  The [\ion{S}{2}]/[\ion{N}{2}] remains fairly constant at about 0.6 for the full range of $z$, with slight increases in the midplane and beyond $z=1$ kpc.

\begin{figure}
\figurenum{3}
\plotone{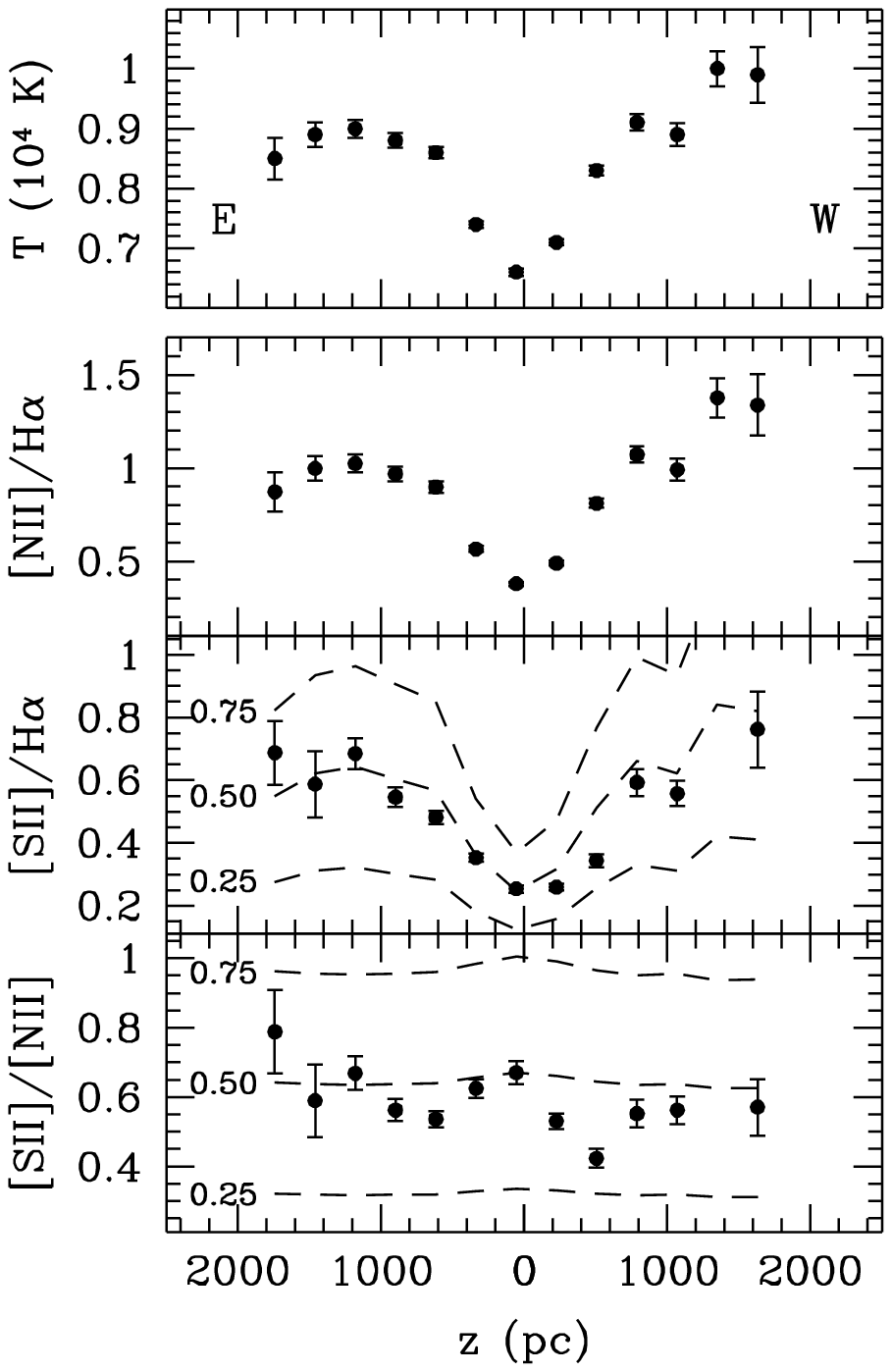}
\caption{Plots of [NII]/H$\alpha$, [SII]/H$\alpha$, and [SII]/[NII] vs. $z$ for NGC 4302. The top plot is the $z$ dependence of the gas temperature determined from the [NII]/H$\alpha$ ratio (see section 3.4).  See caption for Figure 12 for an explanation of symbols, curves, and labels used in the plots.}
\end{figure}

In UGC 10288, the H$\alpha$, [\ion{N}{2}] $\lambda$6583, and [\ion{S}{2}] $\lambda$6716 lines are all detected well above the midplane, with the [\ion{N}{2}] $\lambda$6583 and H$\alpha$ lines being detected up to nearly $z=3$ kpc on the north side for both slit positions.  Detections of the [\ion{O}{3}] $\lambda$5007 and [\ion{O}{1}] $\lambda$6300 lines are limited to $z\lesssim1$ kpc, except for the south side of slit 1 where both lines are detected up to $z=1.5$ kpc.  The vertical runs of [\ion{N}{2}] $\lambda6583/$H$\alpha$, [\ion{S}{2}] $\lambda6716/$H$\alpha$, [\ion{S}{2}] $\lambda6716/$[\ion{N}{2}] $\lambda6583$, [\ion{O}{3}] $\lambda5007/$H$\alpha$, and [\ion{O}{1}] $\lambda6300/$H$\alpha$ for both slits are shown in Figure 4. The data are averaged over 5 spatial pixels.  For slit 1, [\ion{N}{2}]/H$\alpha$ rises from 0.3 at the midplane to 1.6 at $z=2$ kpc on the north side and 1.3 at $z=2$ kpc on the south, while for slit 2 the ratio remains somewhat lower, rising to 1.2 for $z=1.5$ kpc on either side of the disk.  The [\ion{S}{2}]/[\ion{N}{2}] ratio shows a similar range of values for each slit, varying from 0.8 to 1.4 with the lower values generally occurring in the midplane.  These values are significantly higher than those of the other three galaxies.  The [\ion{O}{3}]/H$\alpha$ ratio shows some evidence for an increase with $z$ on the south side of either slit, reaching values as high as 0.75 in Slit 1 and 0.4 in Slit 2.  On the north side of each slit, values reach no higher than 0.5 in Slit 1 and 0.4 in Slit 2.  Such values of [\ion{O}{3}]/H$\alpha$ are consistent with those observed at comparable $z$-heights in NGC 5775.

\begin{figure}
\figurenum{4}
\plotone{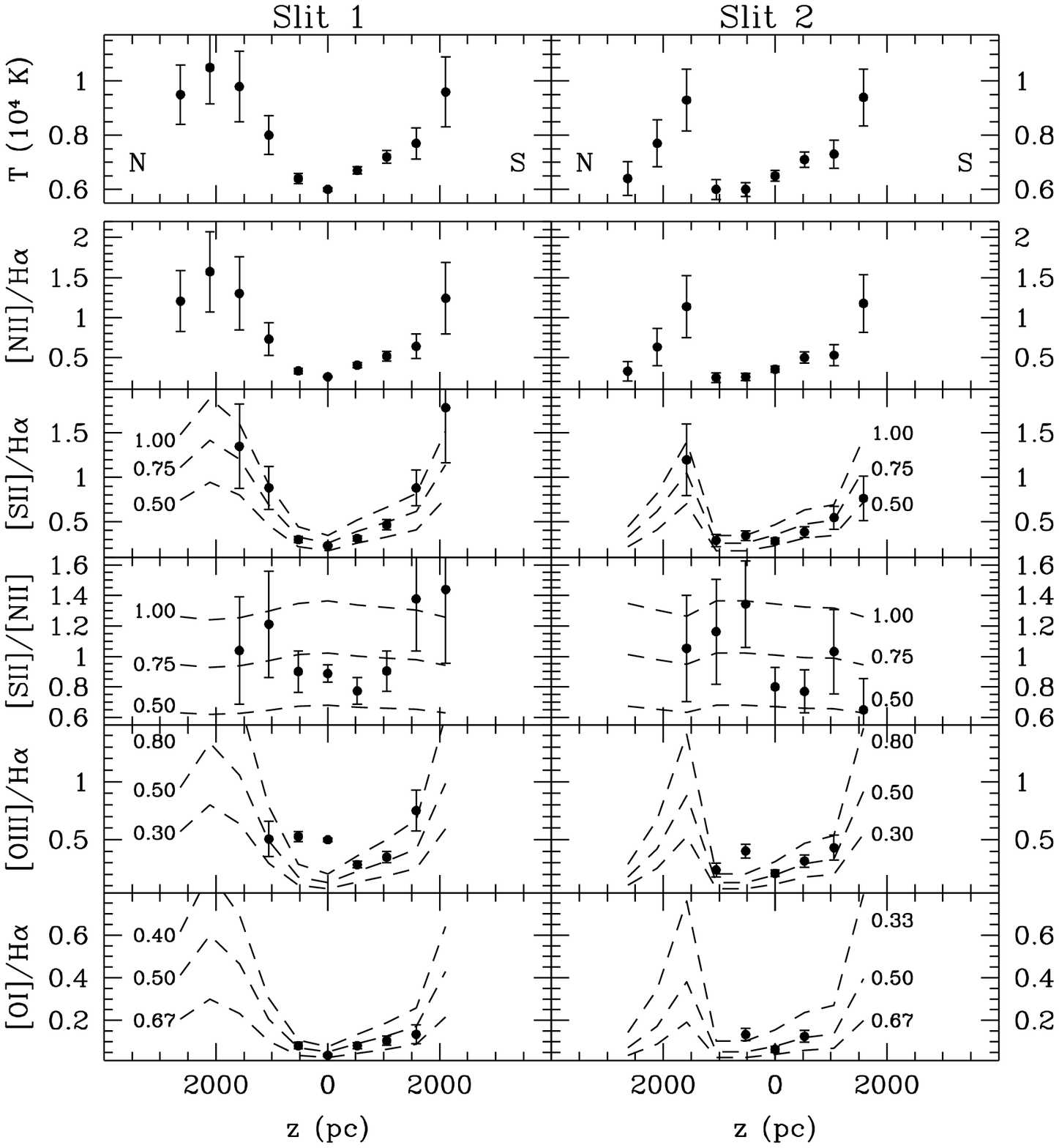}
\caption{Plots of [NII]/H$\alpha$, [SII]/H$\alpha$, [SII]/[NII], [OIII]/H$\alpha$, and [OI]/H$\alpha$ vs. $z$ for both slit positions for UGC 10288. The top plot is the $z$ dependence of the gas temperature determined from the [NII]/H$\alpha$ ratio (see section 3.4).  See caption for Figure 12 for an explanation of symbols, curves, and labels used in the plots.}
\end{figure}

Line ratio data for NGC 5775, except for the [\ion{O}{1}] $\lambda6300/$H$\alpha$ and extinction corrected [\ion{O}{3}] $\lambda5007/$H$\alpha$ ratios,  have been presented previously \markcite{}(Rand 2000).  These, along with [\ion{N}{2}] $\lambda6583/$H$\alpha$, [\ion{S}{2}] $\lambda6716/$H$\alpha$, and [\ion{S}{2}] $\lambda6716/$[\ion{N}{2}] $\lambda6583$ ratios are shown in Figure 12. The data are averaged over 10 spatial pixels.  An increase with $z$ in the extinction corrected [\ion{O}{3}]/H$\alpha$ for the NE side of Slit 1 and the SW side of Slit 2 is still indicated by the data, where values range from 0.2--0.4 in the disk to 1.1 in the halo.  The [\ion{O}{1}/H$\alpha$ ratio ranges from about 0.02 in the midplane to 0.06 at $z=2.5$ kpc in Slit 2, while the [\ion{O}{1}] $\lambda6300$ line was not detected for Slit 1.  Because of the 85\arcdeg\ \markcite{}(Irwin 1994) inclination of the galaxy, points within 10\arcsec\ of the midplane ($z\lesssim1200$ pc in the figures) on the spatial axis reflect in-plane, highly inclined disk structure rather than true vertical structure.  In these lines of sight, \ion{H}{2}\ regions and areas between them have a greater effect on line ratio variations than extraplanar diffuse gas.  This fact explains why some line ratio minima are not at $z=0$ kpc.

Line ratio data for NGC 891, previously presented by Rand \markcite{}(1998), can be seen in Figure 13. 

\subsection{DIG scale heights}
In this section we attempt to determine the H$\alpha$ emission scale height, $H_{em}$, of the galaxies' DIG halos at the various slit locations.  We attempt to model only an exponential halo component, avoiding any disk component where contamination by bright \ion{H}{2}\ regions could complicate the analysis.  We plot the logarithm of H$\alpha$ intensity versus $z$ for each slit, then fit an exponential component with a least-squares fit except in cases where an exponential is obviously a bad description.  Plots of logarithmic H$\alpha$ intensity vs. $z$, along with exponential fits, for NGC 5775, NGC 4302, and UGC 10288 are shown in Figures 5, 6, and 7, respectively.  Results of the fitting procedure are summarized in Table 2.

\begin{figure}
\figurenum{5}
\plotone{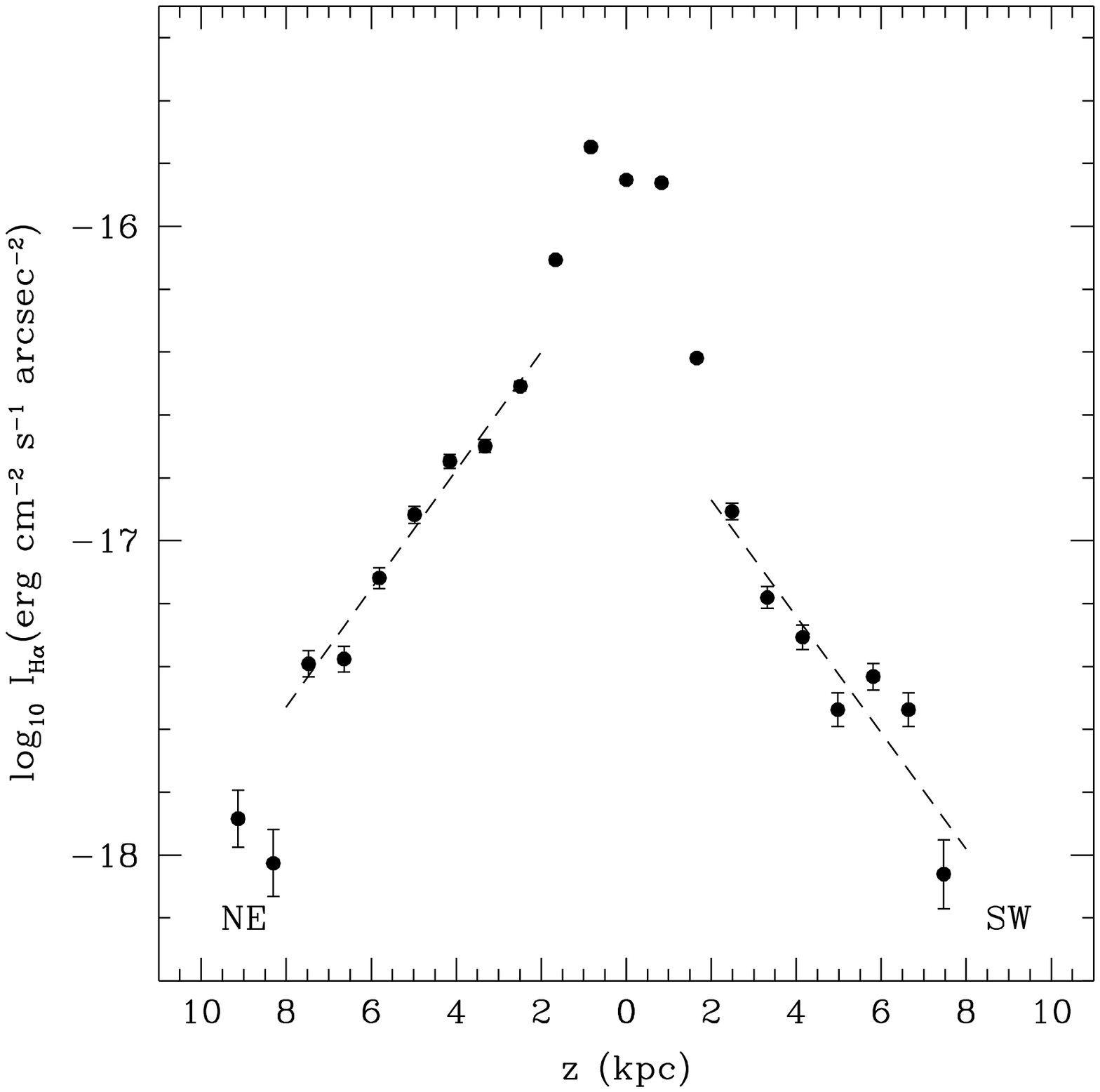}
\caption{Plots of logarithmic H$\alpha$ intensity vs. $z$ (\em{points}\em) for NGC 5775, with exponential least-square fits to the data (\em{dashed lines}\em) for (\em{a}\em) slit position 1 with best fit exponentials of $H_{em}=2300$ pc on each side, and (\em{b}\em) slit position 2 with best fit exponentials of $H_{em}=2500$ pc on the NE side, and $H_{em}=1800$ pc on the SW side.}
\notetoeditor{fig5a.ps and fig5b.ps should both be included within Figure 5.  This is Figure 5a.}
\end{figure}

\begin{figure}
\plotone{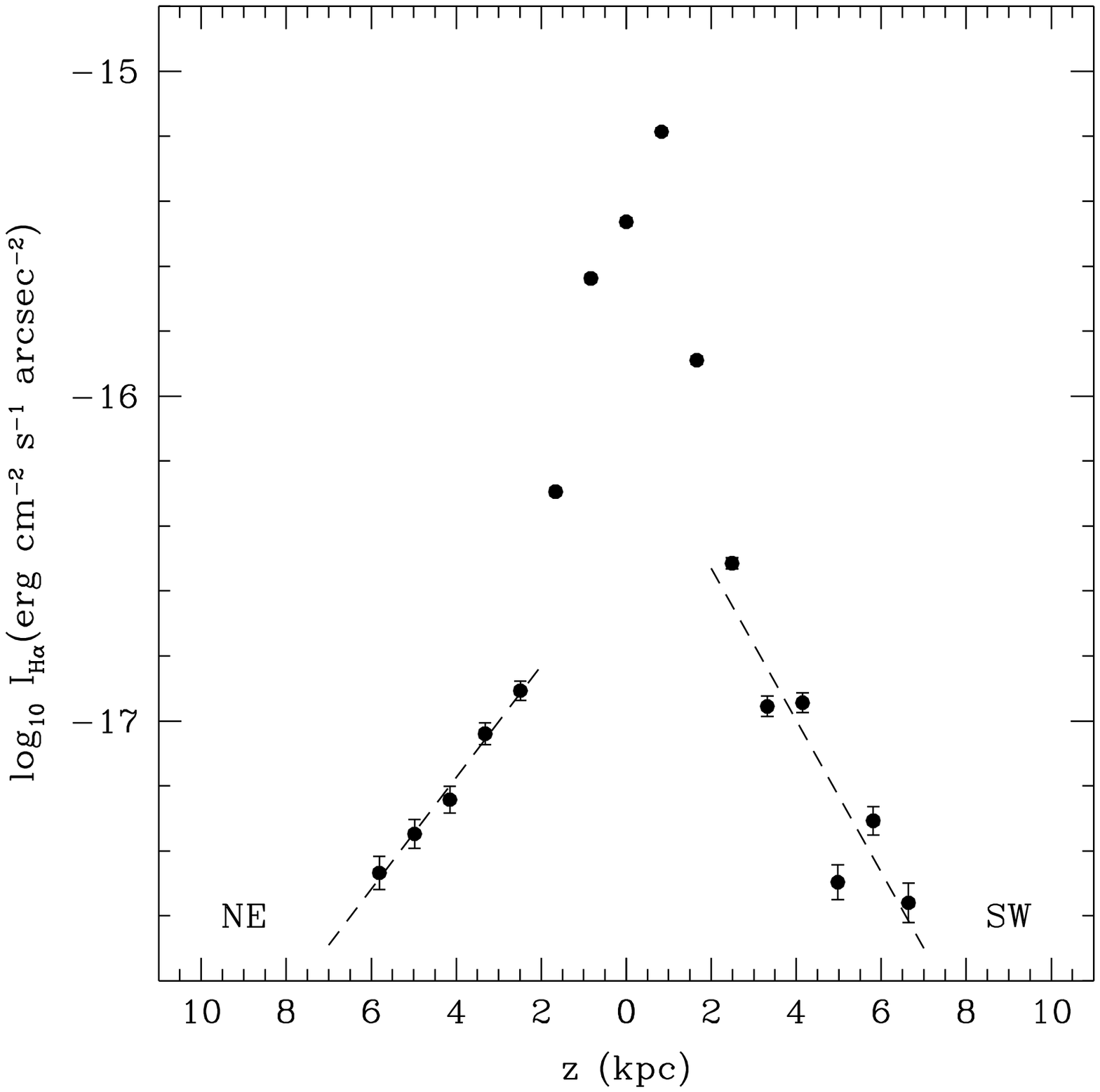}
\notetoeditor{Please include within Figure 5 as Figure 5b.}
\end{figure}

\begin{figure}
\figurenum{6}
\plotone{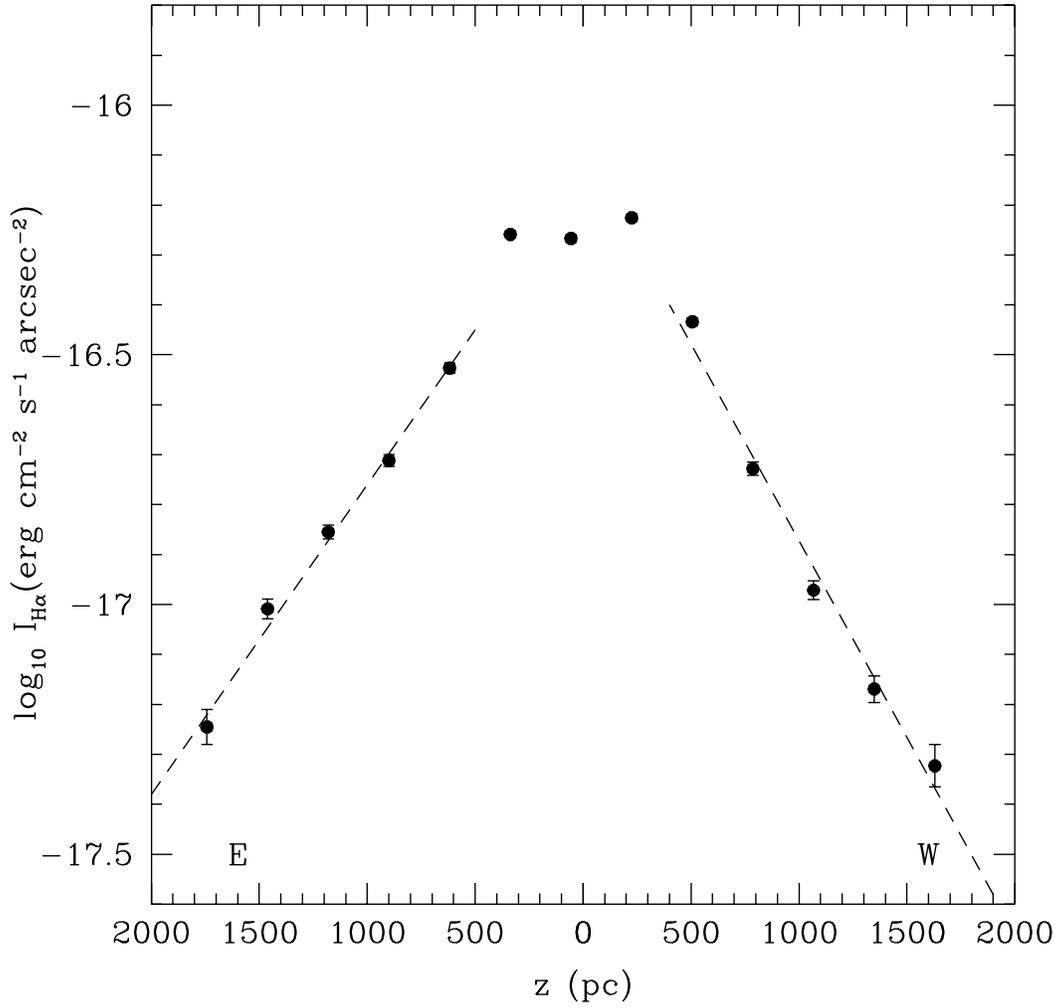}
\caption{Plot of logarithmic H$\alpha$ intensity vs. $z$ (\em{points}\em) for NGC 4302, with exponential least-square fits to the data (\em{dashed lines}\em).  The best fit exponentials to the east and west sides are $H_{em}=700$ pc and $H_{em}=550$ pc, respectively.}
\end{figure}

\begin{figure}
\figurenum{7}
\plotone{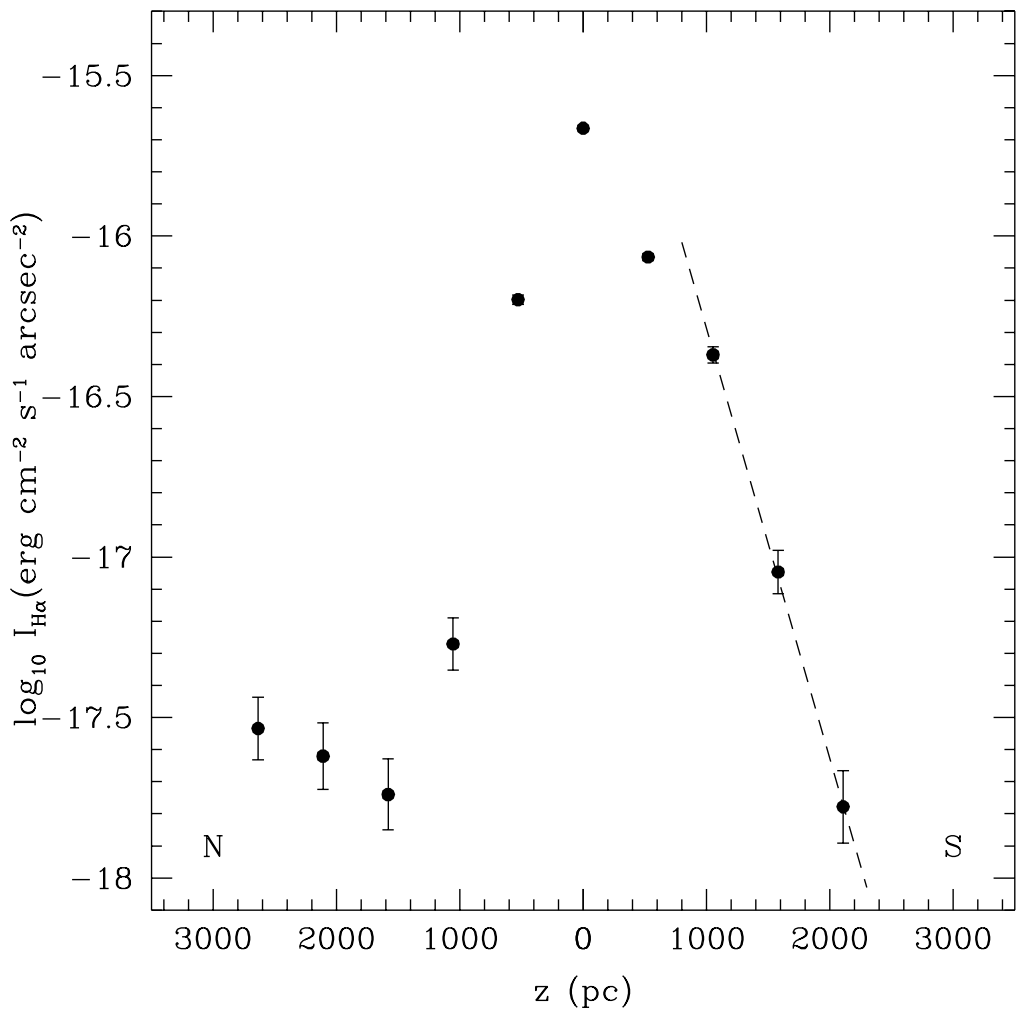}
\caption{Plots of logarithmic H$\alpha$ intensity vs. $z$ (\em{points}\em) for UGC 10288, with exponential least-square fits to the data (\em{dashed lines}\em) for (\em{a}\em) slit position 1 with a best fit exponential of $H_{em}=320$ pc, and (\em{b}\em) slit position 2 with a best fit exponential of $H_{em}=460$ pc.  In both cases only emission on the south side of the disk is well fit by an exponential.}
\notetoeditor{fig7a.ps and fig7b.ps should both be included within Figure 7.  This is Figure 7a.}
\end{figure}

\begin{figure}
\plotone{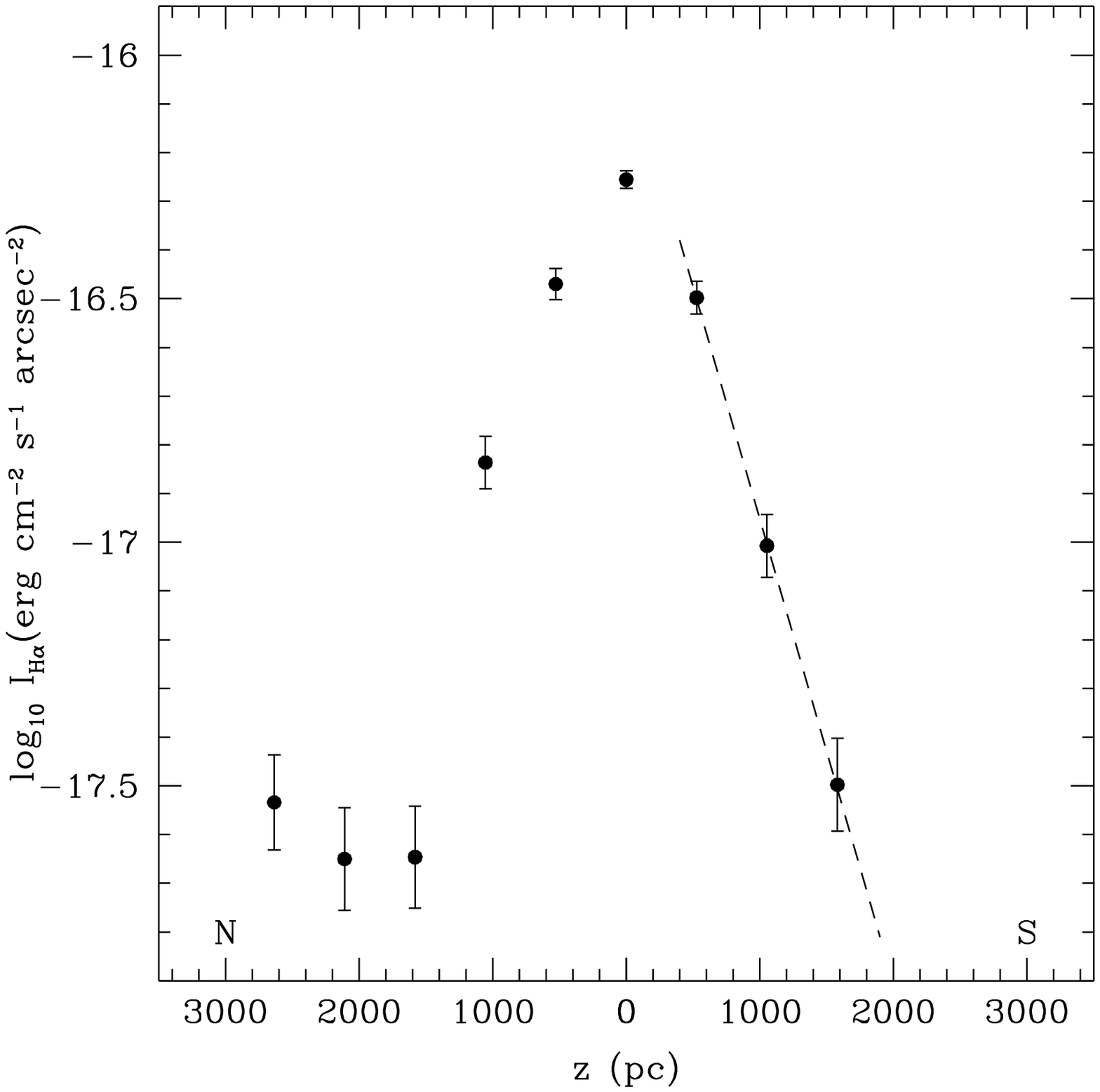}
\notetoeditor{Please include within Figure 7 as Figure 7b.}
\end{figure}

Position 1 of NGC 5775 covers the NE filament, DIG emission from which has been modeled previously \markcite{}(CRDW).  The value $H_{em}=2300$ pc is somewhat consistent with the value of $H=3.0$ kpc determined from image data.  The image data however does include some [\ion{N}{2}] contamination.  The previously determined value assumes an [\ion{N}{2}]/H$\alpha$ ratio that does not vary with $z$. Since our data shows that this ratio in fact rises with $z$ on the NE filament, from 0.5 at $z=2$ kpc to 0.75 at $z=6$ kpc, the value $H=3.0$ kpc becomes an upper limit on the actual scale height.  

The DIG halos of NGC 4302 and UGC 10288 have been modeled previously as well \markcite{}(Rand 1996).  The slit position for NGC 4302 coincides with one of the previously modeled regions, the scale height of which was determined to be $H=525$ pc (the image is not contaminated by [\ion{N}{2}] emission).  The values determined from spectroscopic data, $H_{em}=700$ pc and $H_{em}=550$ pc for the east and west sides, are close to those determined from image data. The scale height determined from the image, however, is an average over a 3 kpc radial extent of the disk and thus should not be in exact agreement with the spectroscopically determined value.  Image data of UGC 10288 revealed very little extraplanar H$\alpha$ emission, and thus profiles were not well modeled with exponentials.  Increased sensitivity of the spectroscopic data have allowed the detection of H$\alpha$ emission to greater $z$-heights.  The resulting H$\alpha$ profiles seem to be well modeled by exponentials with small scale heights on the south side, while more complicated structure is present on the north side.  

\placetable{t2}

NGC 5775, having the most prominent extraplanar DIG emission, also has the DIG halo with the greatest scale height.  In contrast, UGC 10288, showing very little extraplanar H$\alpha$ emission, has a DIG halo of relatively small scale height.  NGC 4302, having a number of extraplanar plumes of H$\alpha$ emission, has a DIG scale height intermediate between that of NGC 5775 and UGC 10288.  The halo of NGC 891 has been modeled previously \markcite{}(RKH), the scale height of which has been measured to be $H_{em}=2.4$ kpc on the west side and $H_{em}=3.2$ kpc on the east side from imaging data.  These results suggest that not only do galaxies such as NGC 5775 and NGC 891 have brighter DIG halos, where sensitivity limits allow faint emission to be detected at much greater heights than for a galaxy with a lower H$\alpha$ surface brightness, but the halos also have much larger scale heights.  This adds further evidence to the notion of a dynamic halo where higher levels of star formation in the disk result in material driven further into the halo (e.g. \markcite{}Rand 1996; \markcite{}Hoopes et. al. 1999).    

\subsection{Shocks as a Secondary Source of Ionization}
Some of the line ratio behavior observed in NGC 5775 and UGC 10288, namely the rise in [\ion{N}{2}]/H$\alpha$ and [\ion{S}{2}]/H$\alpha$ with $z$, is predicted by the photo-ionization models (e.g. S94; Bland-Hawthorn 1997).  As previously stated, rises in these ratios are expected in such models as the radiation field becomes more dilute with $z$.  If the field is in fact becoming more dilute (low values of the ionization parameter, $U$, which measures the ``diluteness''of the radiation field and is proportional to the ratio of ionizing photon number density to gas density) with $z$, then the models predict that the [\ion{O}{3}]/H$\alpha$ ratio should decrease away from the midplane.  However, in NGC 5775, and to a lesser extent UGC 10288, we see a definite incease in [\ion{O}{3}]/H$\alpha$ with $z$, indicative of higher excitation conditions above the disk in such models (high-$U$).  Recent spectroscopy of NGC 5775 by T\"{u}llmann et. al. (2000) also reveals a significant rise in [\ion{O}{3}]/H$\alpha$ with $z$, though these data are not extinction corrected.  It is then possible that the [\ion{O}{3}] emission originates from a different DIG component than that of the [\ion{N}{2}] and [\ion{S}{2}]. Consequently, a secondary mechanism, such as shock ionization, could account for the bright [\ion{O}{3}] emission above the plane.  

In this section we explore the possibility that shock ionization is responsible for some of the DIG emission in these galaxies. One could also consider TMLs, since they also produce large amounts of [\ion{O}{3}] emission.  In fact in NGC 891, TMLs do just as well as shocks in fitting the observed data (Rand 1998).  Our approach is that of Rand \markcite{}(1998), with some minor changes.  We adopt the \markcite{}S94 matter-bounded photoionization model with the lowest terminal hydrogen column density, $1\times10^{18}$ cm$^{-2}$, of individual gas clumps.  The \markcite{}S94 models feature a hard stellar spectrum (an upper mass cutoff of $120 M_{\sun}$), hardening of the radiation field as it propagates through the ISM, and depletion of gas-phase coolants onto dust grains.  We do not consider the S94 radiation-bounded model as this model predicts extremely bright [\ion{O}{3}] emission. Any composite shock model utilizing radiation-bounded models requires a very unusual secondary source of H$\alpha$ emission.  In the case of NGC 5775, such a scenario would require a secondary source that accounts for nearly half of the H$\alpha$ emission at the midplane (assuming log $U\sim-3.0$ at $z=0$ kpc), and produces virtually no [\ion{O}{3}] or [\ion{N}{2}] emission.  In addition, such a source must produce significant [\ion{O}{3}] and [\ion{S}{2}] emission at high-$z$, and still contribute at least 50\% of the total H$\alpha$ emission.  Such emission characteristics could be explained with low-speed shocks near the midplane with a shock velocity increasing with $z$.  However, the problem is further complicated by the fact that low-speed shocks produce copious amounts of [\ion{O}{1}] emission relative to [\ion{N}{2}].  In addition, basic energetic arguments render such a scenario, where the secondary source contributes half of the layer's ionization,  highly unlikely as the ionizing flux from massive stars exceeds supernovae power output, presumably the initiator of these shocks, by a factor of 6 or 7 \markcite{}(Reynolds 1984, 1992).  In any case, parameters describing these shocks require considerable tweaking to account for the observed run of line ratios.  Matter-bounded models on the other hand, better match the data before composite modeling, and require a less complicated parameter space describing the secondary source.

To obtain line ratios due to shock ionization, we use the models of Shull \& McKee \markcite{}(1979; hereafter SM79).  Line ratios in these models, especially [\ion{O}{3}]/H$\alpha$, are highly dependent on shock velocity.  Other variables affecting line ratios include the preshock gas density and ionization state, abundances, and transverse magnetic field strength.  These models assume the gas to be initially neutral at $n_{0}=10$ cm$^{-3}$, then penetrated by a precursor ionization front.  More appropriate to the case of a partially photo-ionized DIG layer, would be a initial density of order $n_{0}\sim0.1$ cm$^{-3}$, and a high initial ionization fraction.  We mainly consider the models with shock velocities of $v_{s}=100$ km s$^{-1}$ and $v_{s}=110$ km s$^{-1}$ in that these higher velocity models are best able to produce large amounts of [\ion{O}{3}] emission, while keeping [\ion{O}{1}] low relative to [\ion{N}{2}].  SM79 also calculate a single model containing depleted abundances with $v_{s}=100$ km s$^{-1}$, which we also consider in the modeling to allow a comparison between shock and photo-ionization models with depleted abundances.  In general, though, we do not consider this analysis as providing a definitive measure of the shock speed in these composite models given the simplicity of the model and the many unrealistic parameters which describe the shocks.   

For the composite modeling, we assume that some fraction of H$\alpha$ emission arises from shock ionization, with that fraction possibly changing with $z$.  We still assume a stellar radiation field distinguished by a decrease in ionization parameter, $U$, with $z$. To simplify matters we assume a single shock velocity for each composite model.  We attempt to fit the composite model to the data by varying the percent contribution at a given value of $U$ (and thus $z$).  For NGC 5775 the modeling worked best by allowing shock contributions to begin at log $U=-4.0$, while models with shocks beginning at log $U=-3.5$ worked best for UGC 10288. Such values are somewhat lower than would be expected for a low-$z$ radiation field.  However, kinematics of edge-on galaxies indicate that at low-$z$, the line emission we observe is mostly associated with the outer disk regions due to an absorbing dust layer. Since star formation is typically concentrated in the inner disk, the radiation field in the outer disk may be relatively dilute, in which case, the value of $U$ at low-$z$ may be quite low (Rand 1998).  

We do not carry out a statistical test of the goodness of fit since our goal is to qualitatively assess the feasibility of a secondary source of ionization being able to account for some of the line ratios.  Although no model reproduces the line ratios to within the errors, we do find that the composite models better reproduce the run of line ratios than photo-ionization models alone.  Rand \markcite{}(1998) finds a similar conclusion for NGC 891, where composite modeling reproduces the run of [\ion{O}{3}]/H$\beta$ vs. [\ion{N}{2}]/H$\alpha$ and [\ion{S}{2}]/H$\alpha$, yet is unable to duplicate the [\ion{S}{2}]/[\ion{N}{2}] ratio at low-$z$.  NGC 4302 is not considered due to the lack of critical [\ion{O}{3}] $\lambda$5007 data.

\subsubsection{NGC 5775}
For the composite modeling, we consider separately the cases of filamentary and non-filamentary emission.  The NE portion of Slit 1 crosses the prominent NE H$\alpha$ filament, previously identified by \markcite{}CRDW as being associated with a shell-like \ion{H}{1} structure and possibly being involved in vigorous disk-halo interactions.  In addition, the SW portion of Slit 2 crosses a bright plume of H$\alpha$ emission immediately adjacent to the very prominent SW H$\alpha$ filament, also a  region where intense disk-halo activity is suspected.  Since these two regions appear associated with current cycling of material from disk to halo, it follows that shock ionization may play a greater role in energizing DIG in these areas.  The SW portion of Slit 1 and the NE portion of Slit 2 each cross regions with relatively less extended DIG emission, where disk-halo interactions appear far less vigorous than regions with more filamentary features.  

The runs of [\ion{O}{3}]/H$\alpha$, [\ion{S}{2}]/H$\alpha$, and [\ion{O}{1}]/H$\alpha$ vs. [\ion{N}{2}]/H$\alpha$ are shown in Figures 8,  9, and 10, respectively. Figures 8 and  9 also present filamentary and non-filamentary regions separately.  The runs of these ratios in the matter-bounded S94 models, with various values of $U$ labeled, are plotted along with predicted line ratios from the SM79 models for $v_{s}=100$ km s$^{-1}$, $v_{s}=110$ km s$^{-1}$, and $v_{s}=100$ km s$^{-1}$ with depletions. We find the $v_{s}=100$ km s$^{-1}$ model is best able to reproduce the observed run of line ratios as a secondary source to photoionization.  The $v_{s}=110$ km s$^{-1}$ model has difficulty reproducing the run of [\ion{S}{2}]/H$\alpha$ vs. [\ion{N}{2}]/H$\alpha$, and the depleted $v_{s}=100$ km s$^{-1}$ model, though able to duplicate the run of [\ion{S}{2}]/H$\alpha$ vs. [\ion{N}{2}]/H$\alpha$, is unable to reproduce the run of [\ion{O}{3}]/H$\alpha$ vs. [\ion{N}{2}]/H$\alpha$ at high-$z$.  

\begin{figure}
\figurenum{8}
\plotone{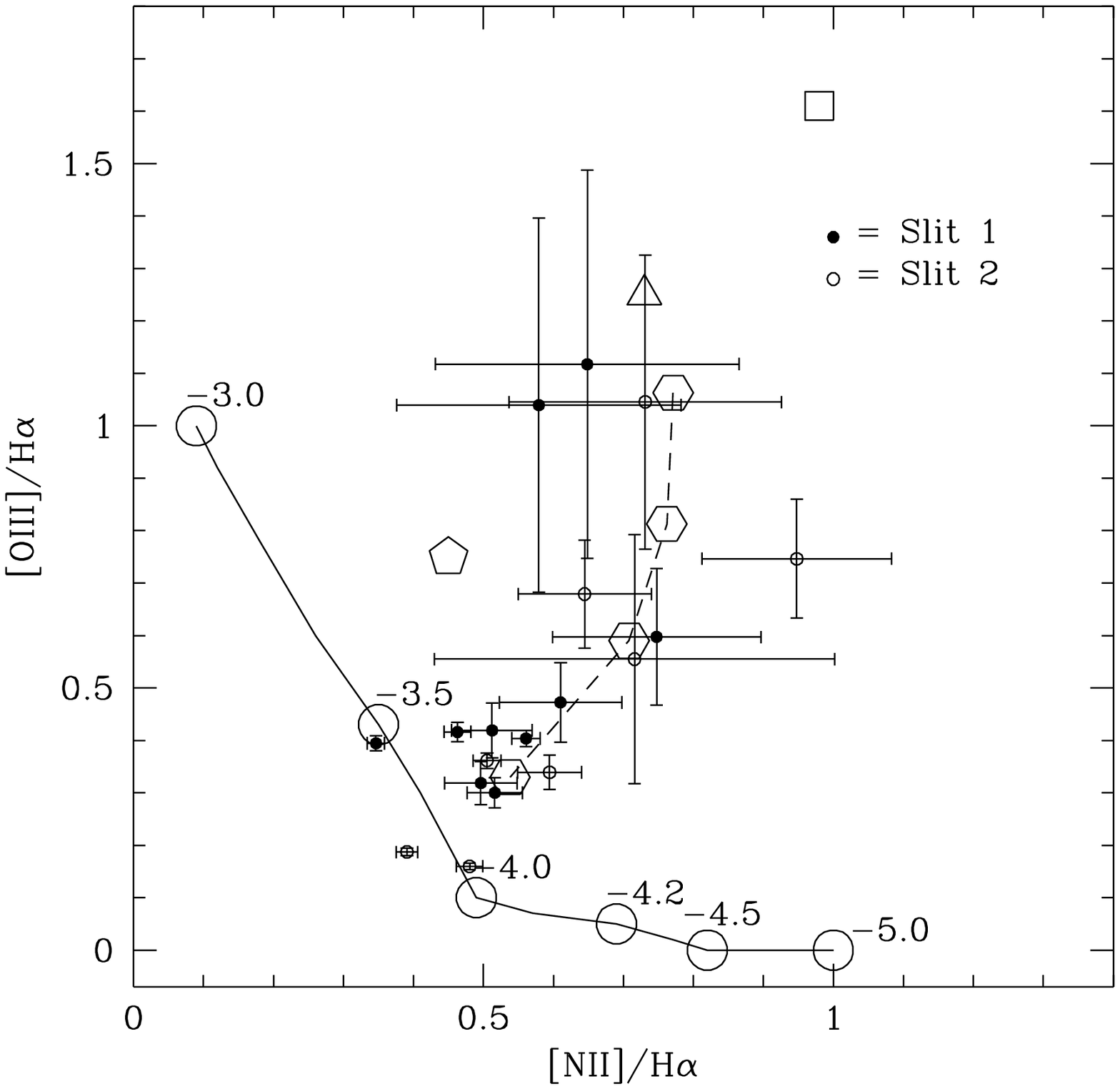}
\caption{Diagnostic diagrams of the line ratios [OIII]/H$\alpha$ vs. [NII]/H$\alpha$ for (\em{a}\em) filamentary regions and (\em{b}\em) non-filamentary regions in NGC 5775.  The solid curve represents predictions from the matter-bounded photo-ionization model of S94, with various values of ionization parameter, log $U$, indicated by circles.  The triangle, square, and pentagon indicate line ratios from SM79 shock models with shock speeds of 100 km/s and 110 km/s with standard abundances, and 100 km/s with depleted abundances, respectively.  The composite photo-ionization/shock model is represented by open hexagons connected by dashed lines.  The hexagons represent ionization parameters for the photo-ionized component of log $U=-4.0, -4.2, -4.5, -5.0$.  The fraction of emission from the 100 km/s shock ranges from 20\% at log $U=-4.0$ to 85\% at log $U=-5.0$ for the filamentary regions and 16\% at log $U=-4.0$ to 27\% at log $U=-5.0$ for the non-filamentary regions.}
\end{figure}

\begin{figure}
\figurenum{8}
\plotone{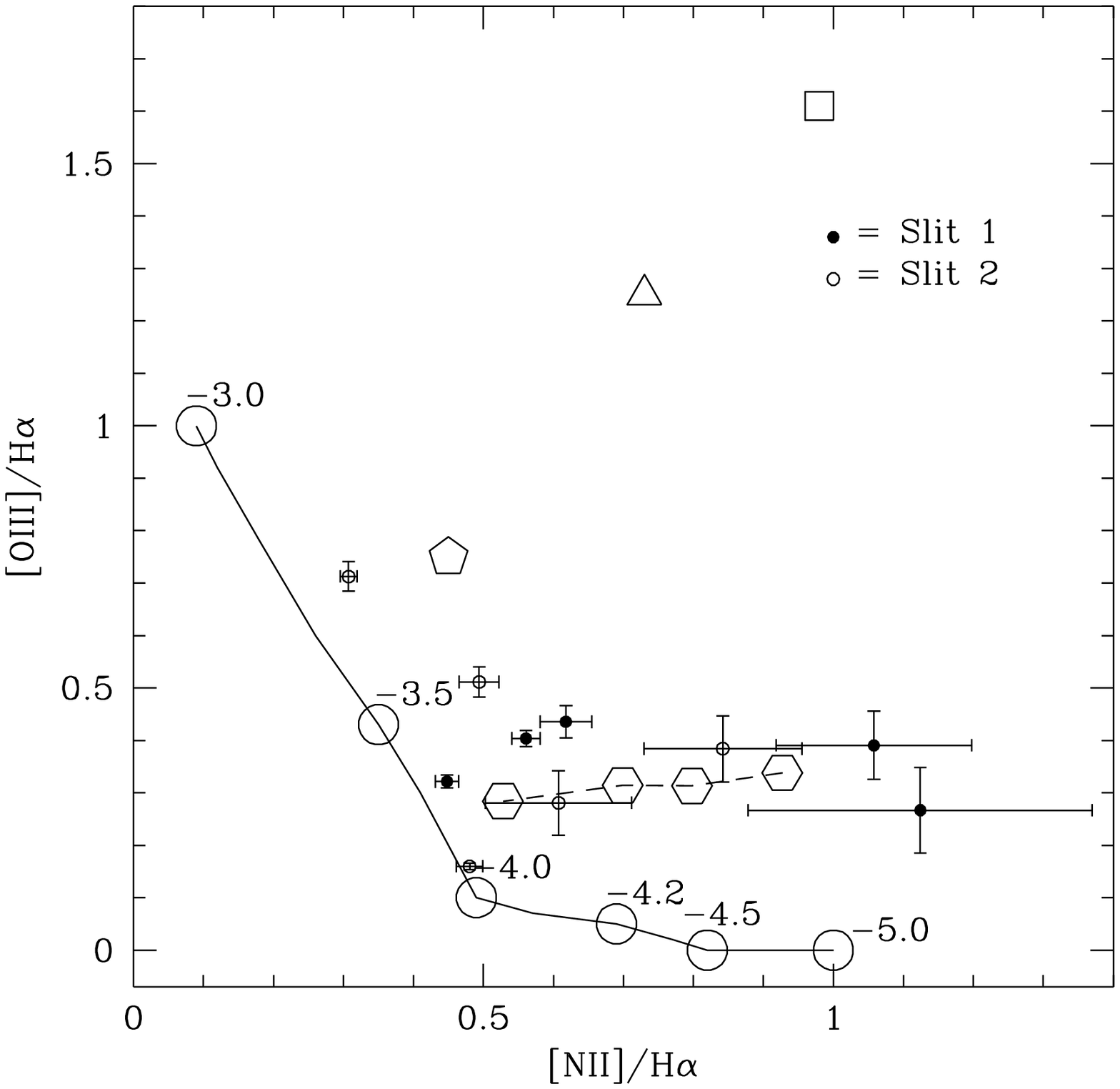}
\notetoeditor{please include within figure 8 as 8b}
\end{figure}

\begin{figure}
\figurenum{9}
\plotone{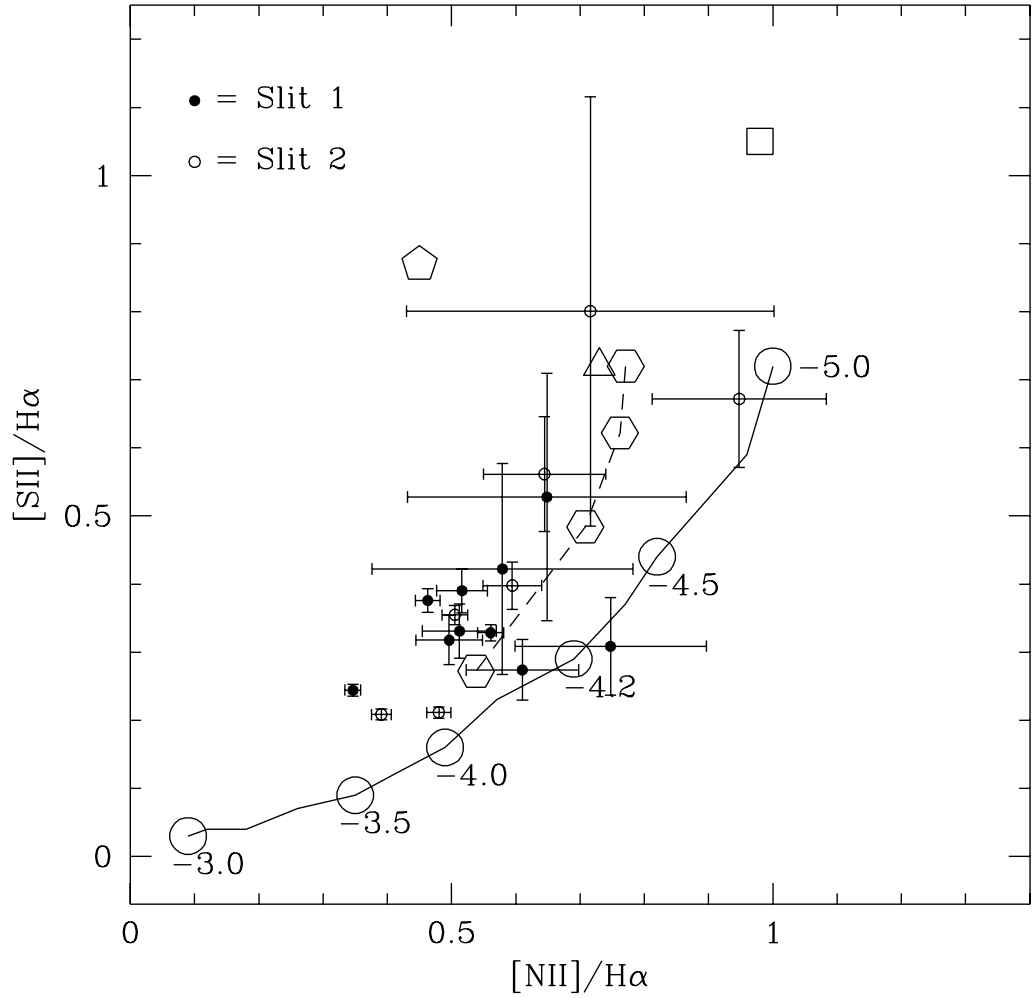}             
\caption{Diagnostic diagrams of the line ratios [SII]/H$\alpha$ vs. [NII]/H$\alpha$ for (\em{a}\em) filamentary regions and (\em{b}\em) non-filamentary regions in NGC 5775.  See the caption for Figure 8 for an explanation of symbols used in the diagram.}
\end{figure}

\begin{figure}
\figurenum{9}
\plotone{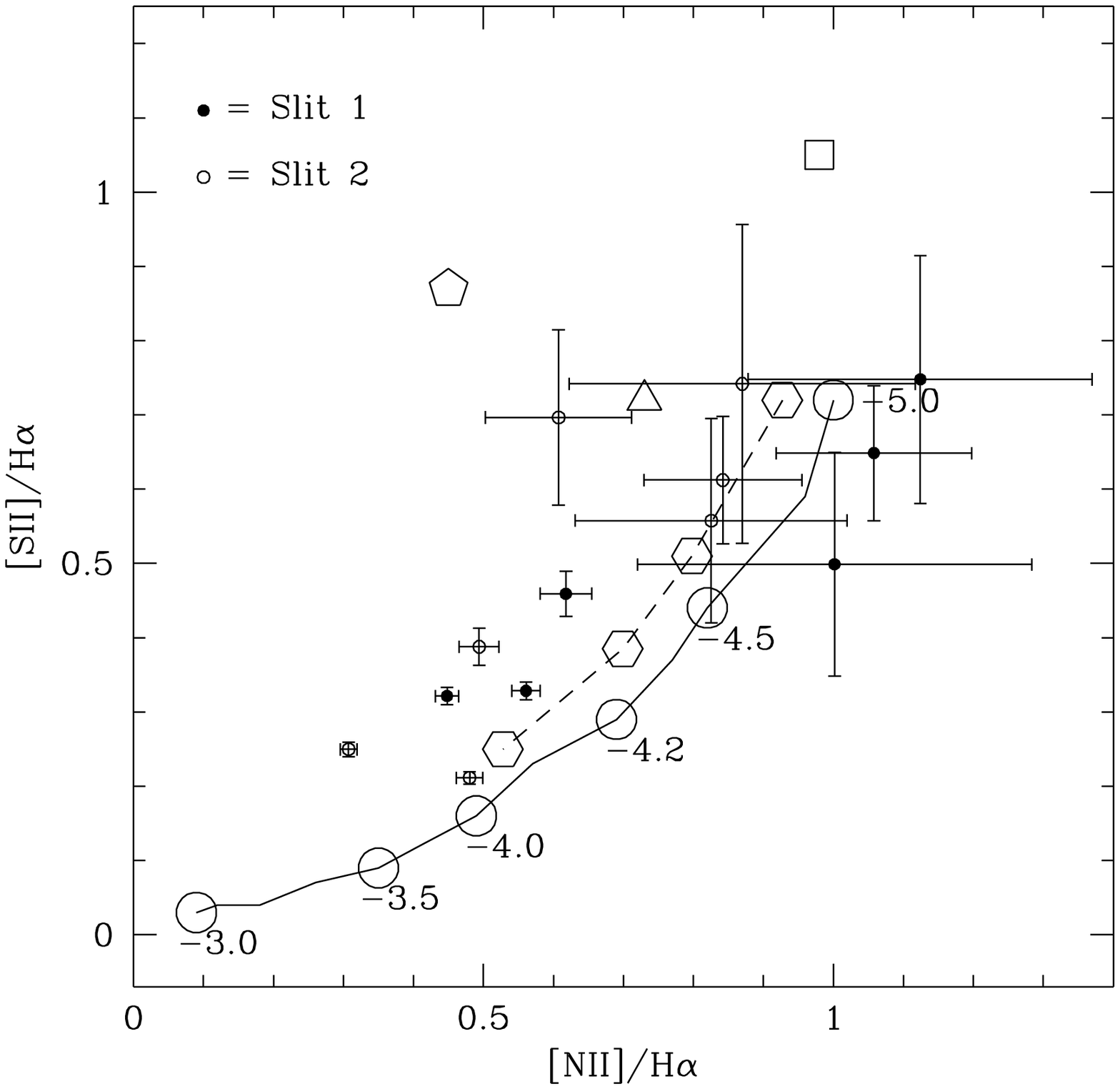}
\notetoeditor{please include within figure 9 as 9b}
\end{figure}

\begin{figure}
\figurenum{10}
\plotone{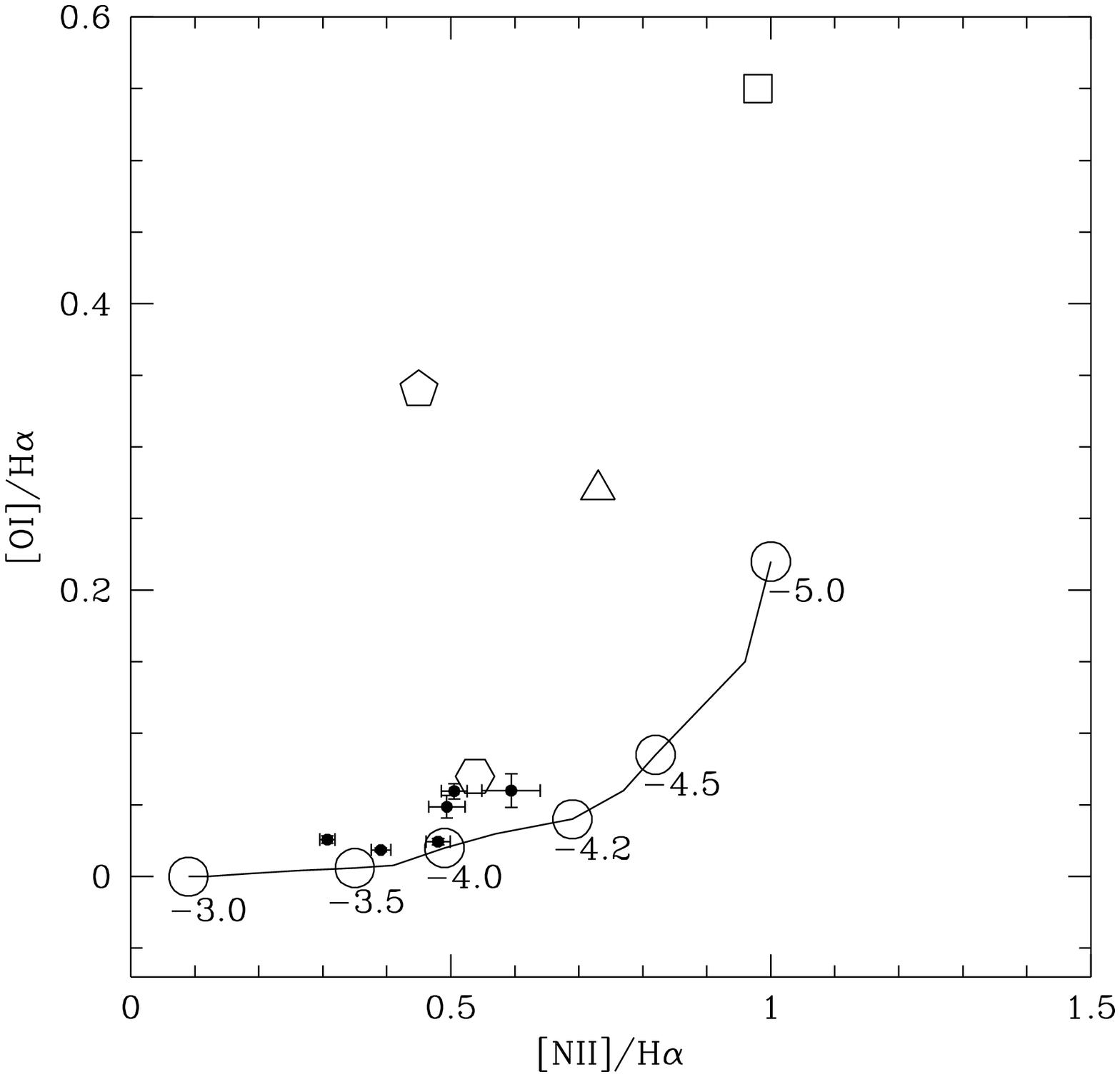}
\caption{Diagnostic diagram of the line ratios [OI]/H$\alpha$ vs. [NII]/H$\alpha$ in NGC 5775 for Slit 2.  Symbols are the same as those described in the caption for Figure 8, except we only plot the composite model at lowest-$z$ (the open hexagon is for log $U=-4.0$).  The composite model shown is the model appropriate for filamentary regions.}
\end{figure} 

We find that the run of [\ion{O}{3}]/H$\alpha$ vs. [\ion{N}{2}]/H$\alpha$ seen in Figure 8a for the filamentary regions is distinctly different from the trend seen in Figure 8b for the non-filamentary regions.  Clearly, for a given value of [\ion{N}{2}]/H$\alpha$, the ratio [\ion{O}{3}]/H$\alpha$ is much greater in DIG with a more filamentary morphology.  The non-filamentary regions show a run that is very similar to the run observed in NGC 891 \markcite{}(Rand 1998). The run of [\ion{S}{2}]/H$\alpha$ vs. [\ion{N}{2}]/H$\alpha$ on the other hand, does not seem to depend strongly on the presence of filamentary DIG structure.  Non-filamentary regions do seem to have greater values of [\ion{N}{2}]/H$\alpha$ and [\ion{S}{2}]/H$\alpha$, in keeping with these ratios' anti-correlation with relative H$\alpha$ surface brightness (see also Figure 12).  All data for the ratio [\ion{O}{1}]/H$\alpha$, being detected only for Slit 2 to $z=2.5$ kpc, are plotted in Figure 10.  The lack of [\ion{O}{1}]/H$\alpha$ data for higher values of [\ion{N}{2}]/H$\alpha$ makes a comparison between various DIG morphologies impossible. 

Composite models are plotted in Figures 8,  9, and 10 as dashed lines joining open hexagons, which mark various values of log $U$.  For the filamentary regions, 20\% of the H$\alpha$ emission at log $U=-4.0$ arises from shock ionization, increasing to 85\% at log $U=-5.0$. This does not mean the majority of extraplanar DIG is shock ionized.  It does indicate however that at very high-$z$ (log $U=-5.0$ corresponds to a height of about $z=6-8$ kpc for the filamentary regions and $z=3-5$ kpc for the non-filamentary), it is possible that most of the emission within the filaments arises from shock ionized gas.  In the non-filamentary case, shocks contribute 16\% of the H$\alpha$  emission at log $U=-4.0$, rising to 27\% at log $U=-5.0$.  For the [\ion{O}{1}]/H$\alpha$ vs. [\ion{N}{2}]/H$\alpha$ data of Figure 10, the filamentary composite model is plotted over all data points for both filamentary and non-filamentary regions.  This model and the non-filamentary model are somewhat successful in explaining the [\ion{O}{1}]/H$\alpha$ vs. [\ion{N}{2}]/H$\alpha$ run at low-$z$, in the sense of matching the excess of [\ion{O}{1}]/H$\alpha$ for a given [\ion{N}{2}]/H$\alpha$ compared to the S94 model alone.

The composite models both match the runs of [\ion{O}{3}]/H$\alpha$ vs. [\ion{N}{2}]/H$\alpha$.  The main deficiency in the models is their inability to reproduce the run of [\ion{S}{2}]/[\ion{N}{2}] at low-$z$.  The models do not predict enough [\ion{S}{2}] emission for a given value of [\ion{N}{2}].  The models perform somewhat better at high-$z$, though the spread in data values at high-$z$ make a trend somewhat difficult to discern.  Again, though models with depleted abundances are better able to produce an excess of [\ion{S}{2}] emission relative to [\ion{N}{2}],  they do not produce enough [\ion{O}{3}] emission.  

In any case, it is clear that the filamentary regions require a greater contribution from shocks to explain the observed line ratios.  The correlation of these features with \ion{H}{1}\ shells and enhanced 20-cm radio continuum emission is also suggestive of a larger percent contribution to the observed emission in these regions from shock ionization.  Enhanced [\ion{O}{3}] emission can also be produced by other sources of ionization such as TMLs.  Rand \markcite{}(1998) found that composite photo-ionization/TML models were just as successful in reproducing the runs of observed line ratios in NGC 891 as composite models featuring shocks.  TMLs are expected to occur at interfaces of gas at different temperatures in the ISM, such as in shell walls.  It is likely then for NGC 5775, where \ion{H}{1}\ shells are associated with the more prominent H$\alpha$ filaments,  that TMLs could also account for the enhanced line ratios.  We have performed cursory modeling of the line ratios for NGC 5775 with composite photo-ionization/TML models.  Using the TML model of Slavin, Shull, \& Begelman \markcite{}(1993) with depleted abundances in the mixing layers of hot and warm gas, a hot gas mixing speed of $v_{t}=25$ km s$^{-1}$, and a mixed gas temperature of log $\bar{T}=5.3$,  we adopt the same approach as the composite photo-ionization/shock modeling.   We find that the photo-ionization/TML models work nearly as well as the photo-ionization/shock models, though they are unable to reproduce the run of ratios for the highest values of [\ion{O}{3}]/H$\alpha$.

\subsubsection{UGC 10288}
A preliminary analysis of the diagnostic diagrams for UGC 10288 has revealed no discernible trend in the run of [\ion{O}{3}]/H$\alpha$ vs. [\ion{N}{2}]/H$\alpha$.  To attempt to better reveal any possible trend, line ratios on either side of the disk have been averaged for each slit to establish a run of each ratio vs. $|z|$.  These averaged runs of [\ion{O}{3}]/H$\alpha$, [\ion{S}{2}]/H$\alpha$, and [\ion{O}{1}]/H$\alpha$ vs. [\ion{N}{2}]/H$\alpha$ are shown in Figure 11.  The matter-bounded S94 models, with various values of $U$ labeled, are plotted along with predicted line ratios from the SM79 models for $v_{s}=100$ km s$^{-1}$, $v_{s}=110$ km s$^{-1}$, and $v_{s}=100$ km s$^{-1}$ with depletions. Even with the averaging, the [\ion{O}{3}]/H$\alpha$ vs. [\ion{N}{2}]/H$\alpha$ run doesn't show an obvious trend: Slit 1 seems to show a slight trend of an  increase in [\ion{O}{3}]/H$\alpha$ with [\ion{N}{2}]/H$\alpha$, while Slit 2 data is clustered near the photo-ionization model.  For this reason, we attempt to fit the run of [\ion{S}{2}]/[\ion{N}{2}] instead of [\ion{O}{3}]/H$\alpha$ vs. [\ion{N}{2}]/H$\alpha$.  We adopt the $v_{s}=100$ km s$^{-1}$ model with depleted abundances as the other two models do not produce enough [\ion{S}{2}] emission relative to [\ion{N}{2}].  

Composite models are plotted in Figure 11 as dashed lines joining open hexagons, which mark various values of log $U$.  We find that the composite models best agree with the data when shocks begin contributing to emission at log $U=-3.5$.  For the composite model, 20\% of the H$\alpha$ emission at log $U=-3.5$ arises from a 100 km s$^{-1}$ shock with depleted abundances, rising to 45\% at log $U=-5.0$ (log $U=-5.0$ corresponds to a height of about $z\approx1000$ pc).

\begin{figure}
\figurenum{11}
\plotone{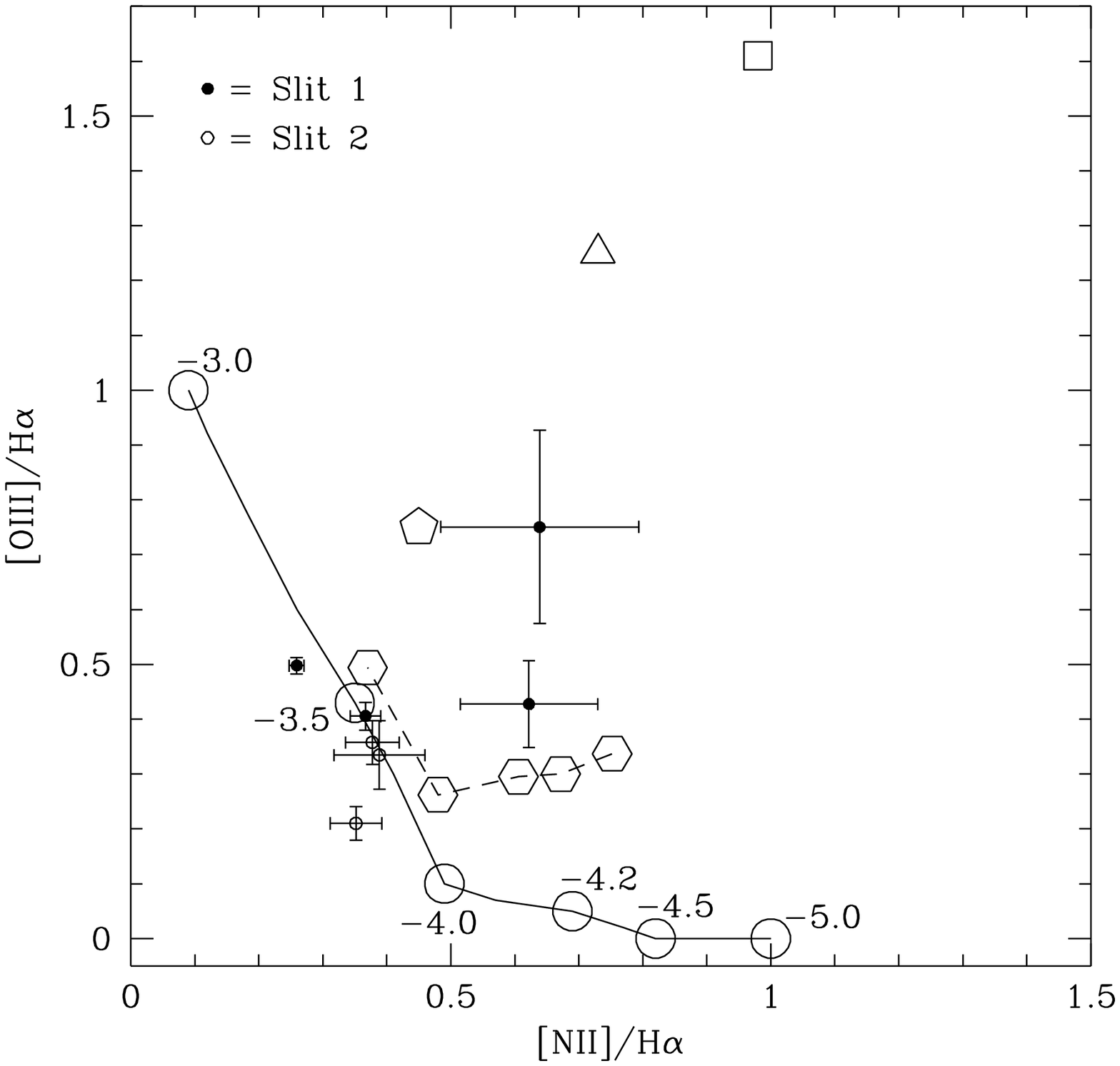}
\caption{Diagnostic diagrams of line ratios for UGC 10288: (\em{a}\em) [OIII]/H$\alpha$ vs. [NII]/H$\alpha$, (\em{b}\em)  [SII]/H$\alpha$ vs. [NII]/H$\alpha$, and (\em{c}\em) [OI]/H$\alpha$ vs. [NII]/H$\alpha$.  Line ratios on either side of the disk have been averaged for each slit.  See Figure 8 for an explanation of symbols used in the diagram.  Open hexagons describing the run of the composite model represent values of ionization parameter, log $U=-3.5, -4.0, -4.2, -4.5, -5.0$, except for the [OI]/H$\alpha$ vs. [NII]/H$\alpha$ composite model which shows only the run at low-$z$ (log $U=-3.5, -4.0. -4.2$).  The fraction of emission from the depleted-abundance 100 km/s model ranges from 20\% at log $U=-3.5$ to to 45\% at log $U=-5.0$.}
\end{figure}

\begin{figure}
\figurenum{11}
\plotone{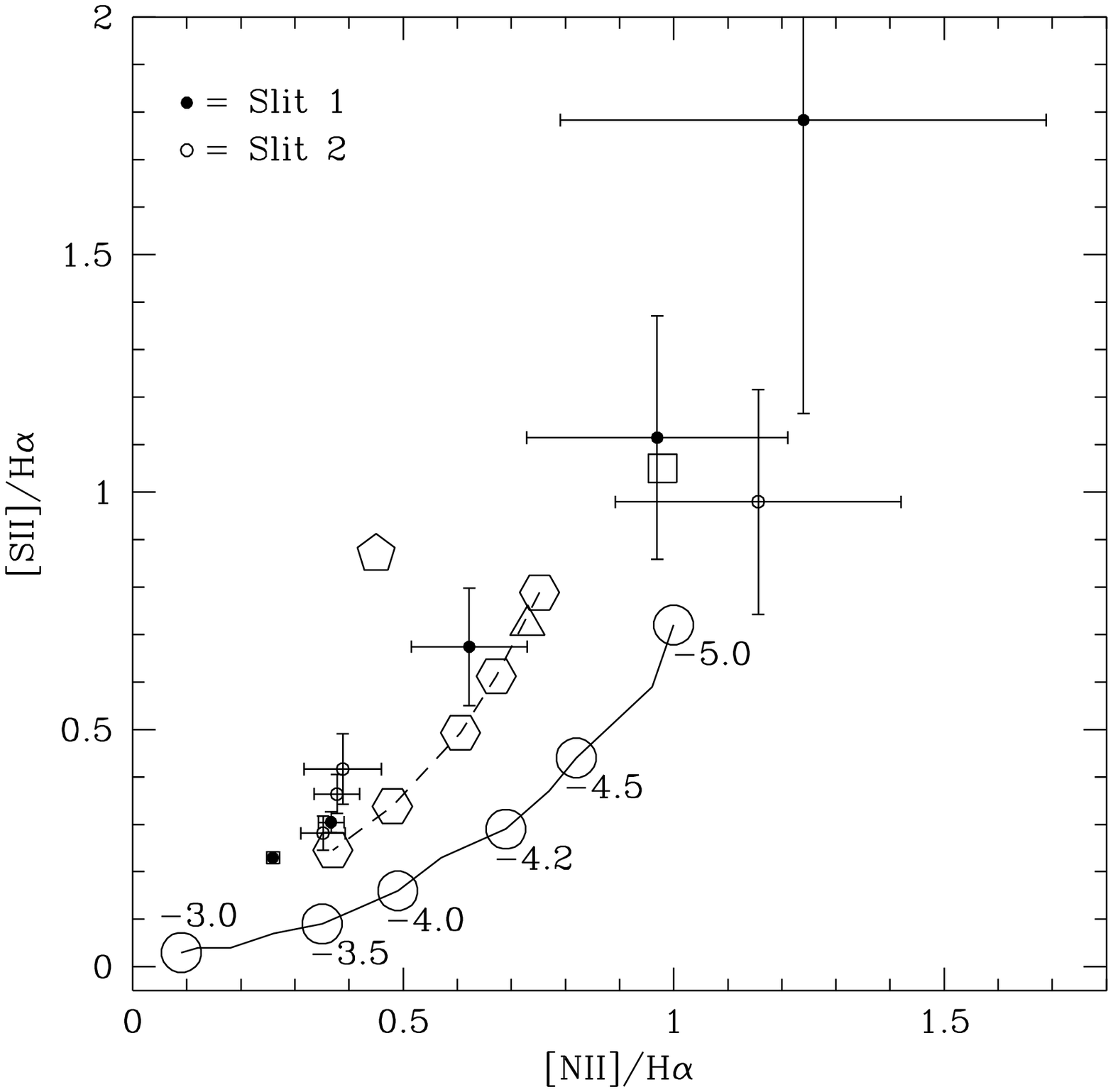}
\notetoeditor{please include within figure 11 as 11b}
\end{figure}

\begin{figure}
\figurenum{11}
\plotone{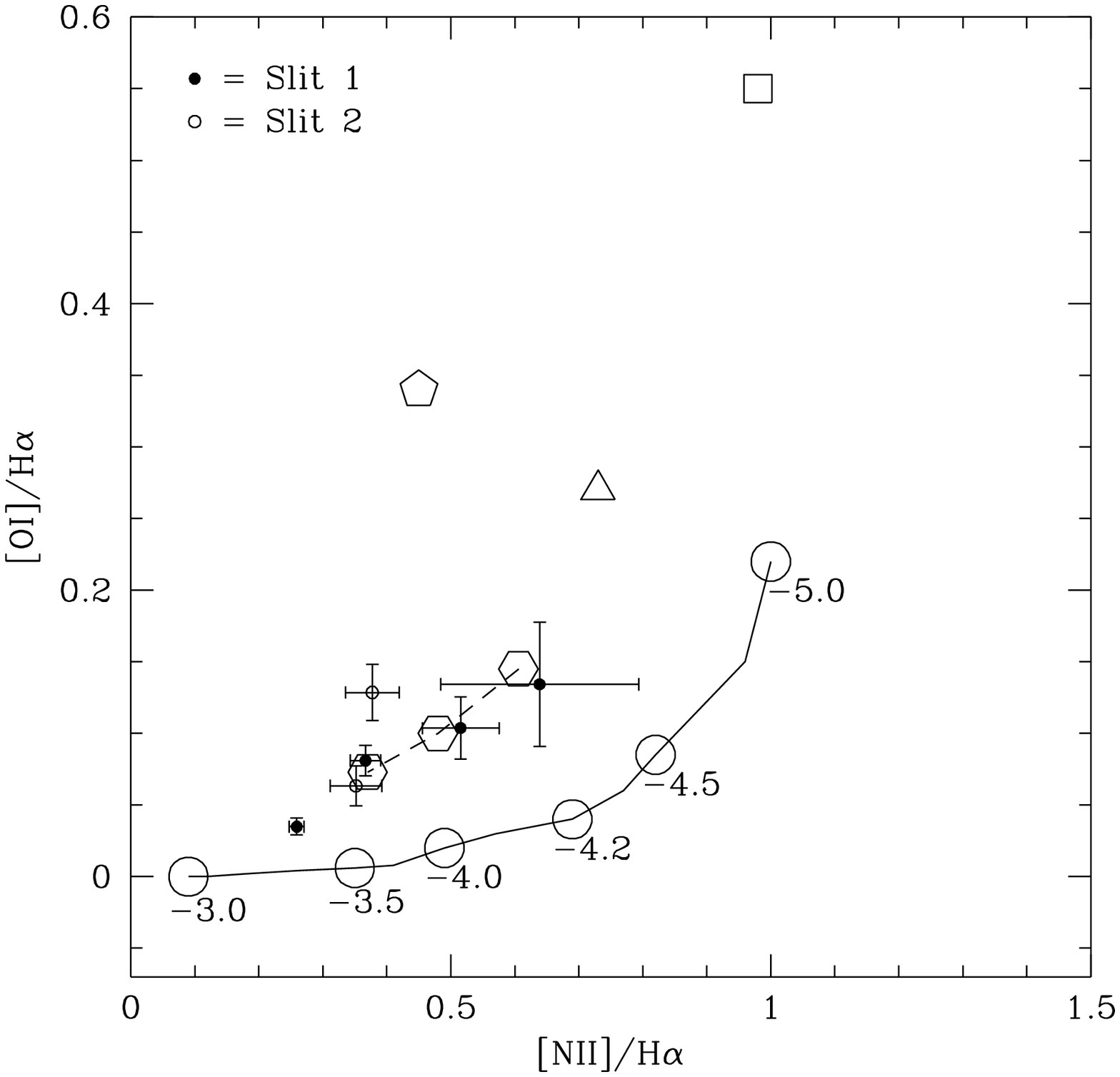}
\notetoeditor{please include within figure 11 as 11c}
\end{figure}   

The composite model successfully replicates the run of [\ion{S}{2}]/[\ion{N}{2}] at all but very high-$z$ where matter bounded photo-ionization models cannot explain values of [\ion{N}{2}]/H$\alpha$ and [\ion{S}{2}]/H$\alpha$ greater than unity.  Radiation bounded photo-ionization models feature greater values of these line ratios; however, predicted runs of [\ion{O}{3}]/H$\alpha$ cannot be reconciled with observations.  The run of [\ion{O}{1}]/H$\alpha$ vs. [\ion{N}{2}]/H$\alpha$ is also well explained by the composite model.  The lack of any obvious trend in the run of [\ion{O}{3}]/H$\alpha$ vs. [\ion{N}{2}]/H$\alpha$ makes it difficult to assess whether shocks are a viable contributor to the H$\alpha$ emission.  However, the high values of [\ion{O}{3}]/H$\alpha$ in Slit 1 do suggest a large contribution from shocks in those regions.  It should be noted, though, that the galaxy has a fairly quiescent appearance in the H$\alpha$ image with little extraplanar emission, calling into question the notion that shocks are permeating the ISM.  It would be useful to obtain more [\ion{O}{3}] data for this galaxy to better diagnose the possible contributions from secondary ionization sources. Nevertheless, a clear departure from a pure photoionization model is indicated by the data.   

\subsection{DIG Temperature as the Main Cause of Line Ratio Variations} 
In this section, following the approach of \markcite{}HRT, we attempt to determine whether variations in line ratios with $z$ can be explained by a change in gas temperature as one moves off the midplane. A temperature increase with $z$ provides a simple explanation of why all ratios of forbidden lines to Balmer lines are generally seen to increase with $z$. In fact, HRT find for the Reynolds Layer that the increase in [\ion{S}{2}]/H$\alpha$ and [\ion{N}{2}]/H$\alpha$, and the constancy of [\ion{S}{2}]/[\ion{N}{2}] with $z$, up to $z=1.75$ kpc, can be accounted for by an increase in gas temperature alone, without having to invoke a secondary source of ionization. Such a scenario requires additional non-ionizing heating for low-density, high-$z$ gas, such as photo-electric heating from dust grains \markcite{}(Reynolds \& Cox 1992) or dissipation of interstellar turbulence \markcite{}(Minter \& Balser 1997). Using line ratio data versus $z$ for NGC 4302, NGC 5775, UGC 10288, and NGC 891, we attempt to test this scenario by determining gas temperatures as well as ionization fractions of constituent elements.

For collisionally excited ions, the equation for intensity of emission lines is given by Osterbrock \markcite{}(1989),
\begin{equation} 
I_{\nu}(\mbox{photons cm$^{-2}$ s$^{-1}$ sr$^{-1}$})=\frac{f_{\nu}}{4\pi} \int n_{i}n_{e} \frac{8.63\times10^{-8}}{T_{4}^{0.5}} \frac{\Omega(i,j)}{\omega_{i}} e^{-(E_{ij}/kT)} dl, 
\end{equation}
where $\Omega(i,j)$ and $E_{ij}$ are the collision strength and energy of the transition, $\omega_{i}$ is the statistical weight of the ground level, $T_{4}$ is the temperature in units of 10$^{4}$ K, and $f_{\nu}$ is the fraction of downward transitions that produce the emission line.  

To simplify matters, we assume no change in physical conditions, namely temperature, density, ionization fraction, and gas abundance, along a given line-of-sight through the DIG layer. Such uniformity of physical parameters within the ISM is clearly not accurate when lines-of-sight cross small scale structures such as \ion{H}{2}\ regions or shocks, where temperature or ionization conditions may change dramatically.  The assumption is valid, however, if these parameters are thought of as averages along a given line of sight instead of describing localized conditions within the ISM. 
 
This assumption, along with the use of the collision strengths provided by Aller \markcite{}(1984), allows the intensity of [\ion{S}{2}], [\ion{N}{2}], and [\ion{O}{3}] emission to be represented, in units of rayleighs, by:

\begin{equation}
I_{6716}(\mbox{R})=2.79\times10^{5}\left(\frac{\mbox{S}}{\mbox{H}}\right) \frac{(\mbox{S$^{+}$}/\mbox{S})}{(\mbox{H$^{+}$}/\mbox{H})} T_{4}^{-0.593} e^{-2.14/T_{4}} EM,
\end{equation}

\begin{equation}
I_{6583}(\mbox{R})=5.95\times10^{4} \left(\frac{\mbox{N}}{\mbox{H}}\right) \frac{(\mbox{N$^{+}$}/\mbox{N})}{(\mbox{H$^{+}$}/\mbox{H})} T_{4}^{-0.474} e^{-2.18/T_{4}} EM,
\end{equation}
and
\begin{equation}
I_{5007}(\mbox{R})=4.84\times10^{4} \left(\frac{\mbox{O}}{\mbox{H}}\right) \frac{(\mbox{O$^{++}$}/\mbox{O})}{(\mbox{H$^{+}$}/\mbox{H})} T_{4}^{-0.38} e^{-2.87/T_{4}} EM.
\end{equation}

The intensity of H$\alpha$ emission is given by,
\begin{equation}
I_{H\alpha}(\mbox{R})=\frac{1}{2.75\ T_{4}^{0.9}} EM,
\end{equation}
where $EM=\int n_{e}^{2} dl$ is the emission measure in units of pc cm$^{-6}$.

A further simplification is to assume N$^{+}$ and H$^{+}$ have approximately equal ionization fractions.  This assumption is justified for two reasons.  First, photoionization models of \markcite{}S94 predict that N$^{+}$/N$^{0}$ tracks H$^{+}$/H$^{0}$ due to similar first ionization potentials (14.5 eV versus 13.6 eV) and a weak charge exchange reaction.  Second, N is not likely to ionize higher than N$^{+}$ in a pure photoionization scenario due to a second ionization potential of 29.6 eV \markcite{}(HRT). A high ionization fraction of O$^{++}$, however, would imply a significant fraction in the N$^{++}$ state as the second ionization potential of oxygen is 35.1 eV. Thus we will be able to check the validity of this assumption with the [\ion{O}{3}]/H$\alpha$ data.  Due to this simplification, of the line ratios available from this data, [\ion{N}{2}]/H$\alpha$ is the best probe of gas temperature. 

The forbidden lines are all dependent on their abundance relative to atomic hydrogen.  The issue of abundances is a complicated one, the understanding of which is critical for interpreting the line ratios.  Current evidence suggests that abundances in halo gas may be quite different from gas in the disk.  Savage and Sembach \markcite{}(1996) find increasing gas-phase abundances in the Milky Way from disk to halo, possibly as a result of partial grain destruction due to shocks.  However, models by Sembach et. al. \markcite{}(2000; hereafter SHRK) predict that the ratios of forbidden lines to Balmer lines tend to rise as abundances fall due to decreased cooling efficiency. For this reason, it is unlikely that abundance variations can explain the increase in ratios of forbidden lines to Balmer lines. However, observations of the [\ion{O}{2}] $\lambda$3727 line, to complement [\ion{O}{3}] $\lambda$5007 and [\ion{O}{1}] $\lambda$6300 measurements,  would be very useful to completely decouple line ratios (which depend on excitation and temperature) from variations in abundance.  To simplify matters, we adopt abundances that do not vary with $z$.  Using galactic gas-phase abundances of S,N, and O of (S/H)$=1.86\times10^{-5}$ \markcite{}(Anders \& Grevesse 1989), (N/H)$=7.5\times10^{-5}$ \markcite{}(Meyer, Cardelli, \& Sofia 1997), and (O/H)$=3\times10^{-4}$ \markcite{}(Cardelli \& Meyer 1997), along with the assumption of equal fractions of N$^{+}$ and H$^{+}$, we arrive at expressions relating line ratios to temperature and ionization fraction:
\begin{equation}
\frac{[\mbox{\ion{N}{2}}]}{\mbox{H$\alpha$}}=12.2\ T_{4}^{0.426} e^{-2.18/T_{4}},
\end{equation}

\begin{equation}
\frac{[\mbox{\ion{S}{2}}]}{\mbox{H$\alpha$}}=14.3 \left(\frac{\mbox{S$^{+}$}/\mbox{S}}{\mbox{H$^{+}$}/\mbox{H}}\right) T_{4}^{0.307} e^{-2.14/T_{4}},
\end{equation}

\begin{equation}
\frac{[\mbox{\ion{S}{2}}]}{[\mbox{\ion{N}{2}}]}=1.2 \left(\frac{\mbox{S$^{+}$}/\mbox{S}}{\mbox{H$^{+}$}/\mbox{H}}\right) T_{4}^{-0.119} e^{0.04/T_{4}},
\end{equation}
and
\begin{equation}
\frac{[\mbox{\ion{O}{3}}]}{\mbox{H$\alpha$}}=40 \left(\frac{\mbox{O$^{++}$}/\mbox{O}}{\mbox{H$^{+}$}/\mbox{H}}\right) T_{4}^{0.52} e^{-2.87/T_{4}}.
\end{equation}

The ionization fractions of oxygen and hydrogen are coupled through a charge exchange reaction that allows the ionization state of Hydrogen to be determined from the [\ion{O}{1}]/H$\alpha$ ratio. We use the ratio of emissivities of the [\ion{O}{1}] $\lambda$6300 line to the H$\alpha$ line provided by Reynolds et. al. \markcite{}(1998a) to obtain, 
\begin{equation}
\frac{[\mbox{\ion{O}{1}}]}{\mbox{H$\alpha$}}=7.9 \left[\frac{1-\left(\frac{\mbox{H$^{+}$}}{\mbox{H}}\right)}{\left(\frac{\mbox{H$^{+}$}}{\mbox{H}}\right)}\right] \frac{T_{4}^{1.85}}{1+0.605T_{4}^{1.105}} e^{-2.284/T_{4}}.
\end{equation}

Gas temperatures are derived from the [\ion{N}{2}]/H$\alpha$ ratio.  Assuming this ratio accurately reflects temperatures within the medium, line ratios involving the S and O lines are then calculated for a variety of ionization fractions. We can then see whether all the ionization fractions are reasonable and consistent with our initial assumptions. Plots of temperature versus $z$ and calculated ratios overlayed on measured ratios are shown in Figures 3, 4, 12, and 13 for NGC 4302, UGC 10288, NGC 5775, and NGC 891, respectively.  Resulting ionization fractions from the temperature modeling are shown in Table 3.   

\begin{figure}
\figurenum{12}
\plotone{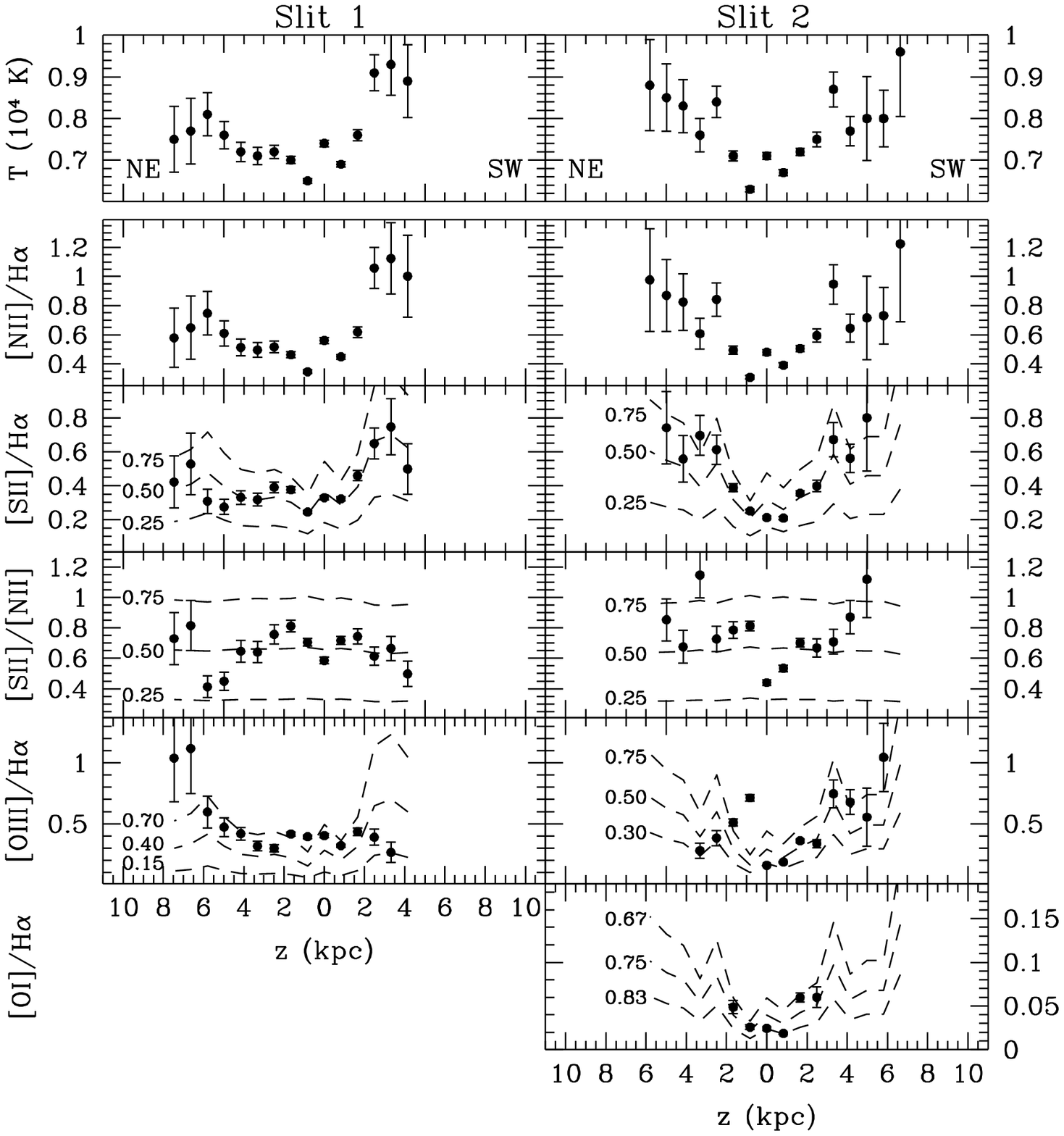}
\caption{Plots of [NII]/H$\alpha$, [SII]/H$\alpha$, [SII]/[NII], [OIII]/H$\alpha$, and [OI]/H$\alpha$ vs. $z$ for both slit positions for NGC 5775.  The top plot is the $z$ dependence of the gas temperature determined from the [NII]/H$\alpha$ ratio.  Dashed curves in the line ratio plots are ratios calculated from the gas temperature for various ionization fractions.  The relevant ionization fractions are:  $\left(\frac{\mbox{S$^{+}$/S}}{\mbox{H$^{+}$/H}}\right)$ for [SII]/H$\alpha$ and [SII]/[NII], $\left(\frac{\mbox{O$^{++}$/O}}{\mbox{H$^{+}$/H}}\right)$ for [OIII]/H$\alpha$, and $\left(\frac{\mbox{H$^{+}$}}{\mbox{H}}\right)$ for [OI]/H$\alpha$.  Numbers adjacent to the dashed curves in the line ratio plots indicate the value of the relevant ionization fraction.}
\end{figure}

\begin{figure}
\figurenum{13}
\plotone{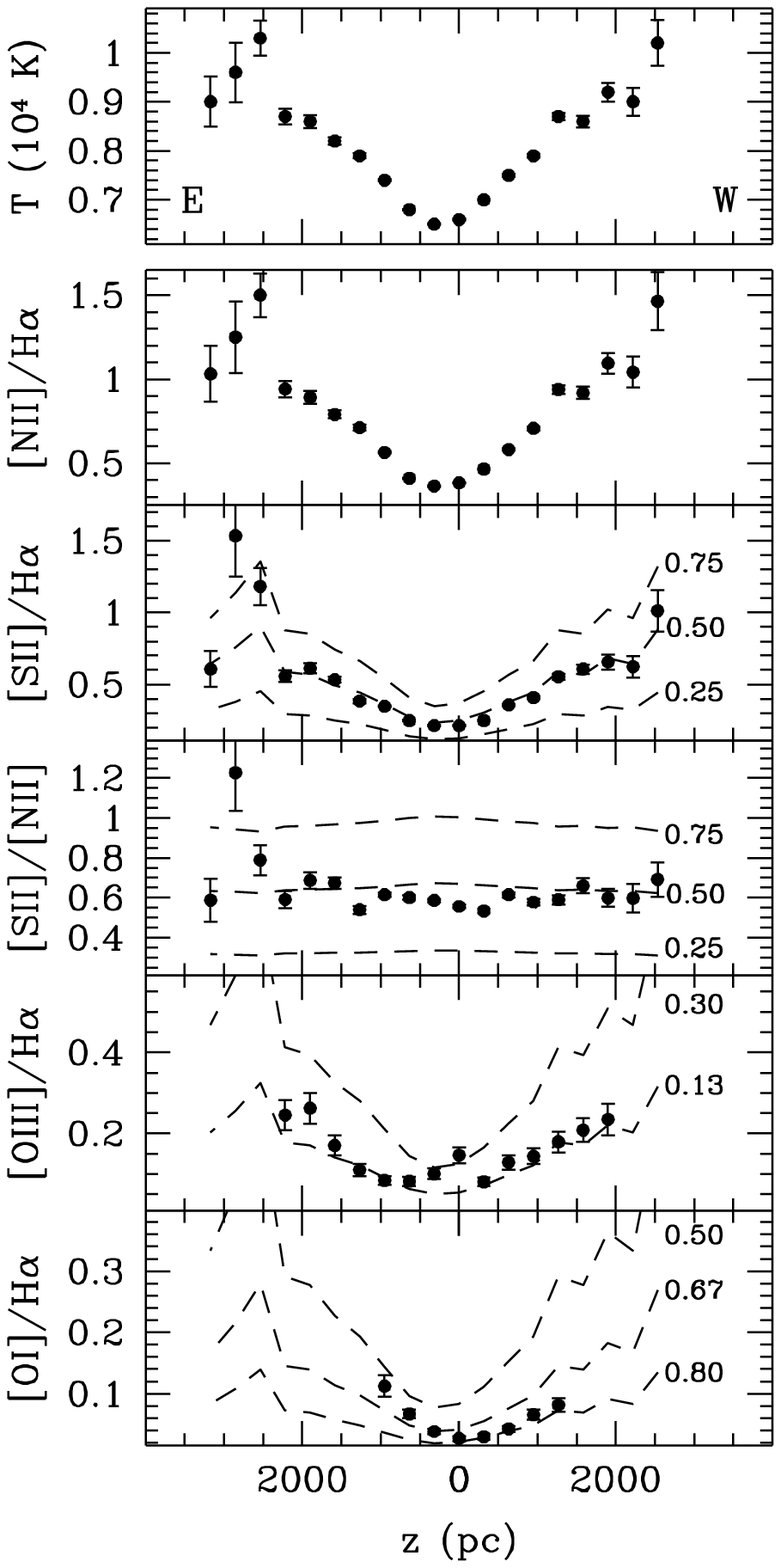}
\caption{Plots of [NII]/H$\alpha$, [SII]/H$\alpha$, [SII]/[NII], [OIII]/H$\alpha$, and [OI]/H$\alpha$ vs. $z$ for NGC 891. The top plot is the $z$ dependence of the gas temperature determined from the [NII]/H$\alpha$ ratio.  See caption for Figure 12 for an explanation of symbols, curves, and labels used in the plots.}
\end{figure}
  
\placetable{t3}

To first order, a rise in DIG temperature can account for the observed rises in line intensities relative to H$\alpha$ emission, as well as the near constancy in the [\ion{S}{2}]/[\ion{N}{2}] ratio.  We find, except in the case of UGC 10288, the ionization fraction of S$^{+}$ relative to H$^{+}$ remains very close to 0.5 independent of $z$.  This is similar to the result obtained by \markcite{}HRT for the WIM.  Photo-ionization models performed by \markcite{}SHRK predict a comparable S$^{+}$ ionization fraction for low-density \ion{H}{2}\ regions.  UGC 10288 on the other hand, shows a considerably higher fraction of S$^{+}$ relative to H$^{+}$ as well as some evidence for an increase with $z$.  However, the ionization fraction of H$^{+}$, determined from points relatively close to the midplane, is significantly lower than in the case of NGC 5775 or NGC 891.  As a result, the range in ionization fraction of S$^{+}$ in UGC 10288 is nearly equal to the range in values for the other galaxies. This does not explain the observed rise with $z$, however, and the lack of [\ion{O}{1}] data at high-$z$ makes it impossible to determine whether the ionization state of sulfur changes with $z$.

The [\ion{O}{3}]/H$\alpha$ data, in addition to revealing information about the ionization state of O, is also an excellent test of the validity of some of the assumptions used for this temperature modeling.  The data for NGC 5775 and UGC 10288 suggest that a significant fraction of oxygen is in the O$^{++}$ state; at least 15\% and in some lines-of-sight nearly 100\% relative to H$^{+}$.  The second ionization potential of oxygen is 35.1 eV, while that of nitrogen is 29.6 eV.  Thus one would expect the ionization fractions of the doubly-ionized species of these two elements to roughly track one another in a warm medium.  The determined values of ionization fraction of O$^{++}$ for these two galaxies imply that a significant percentage of nitrogen atoms are in the N$^{++}$ state, particularly at high-$z$ for NGC 5775.  The likely presence of significant amounts of N$^{++}$ as well as the possibility of its fraction changing with $z$ complicates the use of [\ion{N}{2}]/H$\alpha$ as a temperature indicator.  In any event, the high values of O$^{++}$ ionization fraction suggest that a temperature increase alone is not able to fully explain the runs of line ratios observed in NGC 5775 and UGC 10288 since pure photo-ionization models predict very low O$^{++}$ fractions.  One possible explanation for the large ionization fractions of doubly-ionized species is that a secondary ionizing source with emission characteristics similar to shocks or TMLs contributes to the layer's ionization.  Alternatively,  the [\ion{O}{3}] emitting gas could represent a component that is at a significantly higher temperature than the component responsible for the [\ion{N}{2}] emission. 

NGC 891 on the other hand, shows considerably lower values for the ionization fraction of O$^{++}$ relative to H$^{+}$: 13\% on either side of the disk and just above 30\% in the midplane.  The slit runs near a bright filament that is not as prominent as those observed in NGC 5775. These fractions are slightly larger than those predicted by the SHRK model for low-density \ion{H}{2}\ regions, though still more than an order of magnitude higher than that of the diffuse gas model of SHRK.  Thus it seems the run of line-ratios in NGC 891 is better explained by an increase in gas temperature with $z$ than for NGC 5775.  A slightly warmer [\ion{O}{3}] emitting component than indicated by the [\ion{N}{2}]/H$\alpha$ ratio could explain measured values of [\ion{O}{3}]/H$\alpha$ while keeping the ionization fraction of O$^{++}$ at levels predicted by the models.  If this is the case, a secondary ionizing source would not be necessary to explain the runs of line ratios in NGC 891.   

The run of [SII]/H$\alpha$ in NGC 4302 is well fit by the increase in temperature reflected by the [NII]/H$\alpha$ ratio.  It would be interesting to obtain [OIII]$\lambda$5007 data for this galaxy to better assess whether line ratio variations can be explained by DIG temperature variations.

Other factors may influence the line ratios such as afore-mentioned abundance gradients \markcite{}(Savage \& Sembach 1996) as well as scattered light due to extraplanar dust.  The effects of dust absorption on extraplanar emission have been shown to be significant for a number of DIG halos \markcite{}(Howk \& Savage 1999).  Modeling of the Milky Way \markcite{}(Wood \& Reynolds 1999) and NGC 891 \markcite{}(Ferrara et. al. 1996) suggest that scattered light makes a minor contribution to diffuse halo H$\alpha$ emission that decreases with $z$. In NGC 891, the contribution to DIG emission from scattered light falls to 10\% at $z=600$ pc \markcite{}(Ferrara et. al. 1996).  In any case, dust-scattered light may be an important consideration for the interpretation of these line ratios.   

\section{Conclusions}
Deep spectroscopy of four edge-on spiral galaxies has allowed physical conditions within their DIG halos to be probed, and has provided information on possible sources of ionization of the gas.  This study has revealed the following conclusions:

1. The general trend for the runs of line ratios for NGC 4302 and UGC 10288 is very similar to that of NGC 5775 \markcite{}(Rand 2000) and NGC 891 \markcite{}(Rand 1998): an increase in [\ion{N}{2}]/H$\alpha$ and [\ion{S}{2}]/H$\alpha$ with [\ion{S}{2}]/[\ion{N}{2}] remaining nearly constant with $z$, though one slit position for UGC 10288 does show evidence for an increase in [\ion{S}{2}]/[\ion{N}{2}] with $z$.  Determinations of halo emission scale height have revealed that the more prominent DIG halos also have greater scale height.  This is consistent with models of dynamic halos where greater levels of star formation in the disk can push material farther off the midplane.

2. Pure photo-ionization models do not explain the runs of line ratios in NGC 5775 and UGC 10288.  Composite photo-ionization/shock models can replicate the run of [\ion{O}{3}]/H$\alpha$ vs. [\ion{N}{2}]/H$\alpha$ in NGC 5775, though the run of [\ion{S}{2}]/H$\alpha$ vs. [\ion{N}{2}]/H$\alpha$ is still problematic, especially at low-$z$.  In addition, we find that the composite models require a greater percent contribution from shocks in filamentary than non-filamentary regions to match the run of [\ion{O}{3}]/H$\alpha$.  A lack of a discernible trend for the [\ion{O}{3}]/H$\alpha$ vs. [\ion{N}{2}]/H$\alpha$ run in UGC 10288 makes it difficult to determine whether shocks contribute to line emission, though the high values of [\ion{O}{3}]/H$\alpha$ in Slit 1 indicate that a shock contribution may be necessary.  A composite model, though, can account for the run of [\ion{S}{2}]/H$\alpha$ vs. [\ion{N}{2}]/H$\alpha$.  However, if it follows the trend seen in NGC 5775 of more disturbed regions requiring a larger shock contribution, then the lack of a bright, filamentary halo in UGC 10288 may imply that shocks do not play as significant a role in layer's ionization.

The [\ion{O}{3}]/H$\alpha$ vs. [\ion{N}{2}]/H$\alpha$ run seems to be a key discriminant for shocks.  Fabry-Perot observations of these 3 lines for brighter halos, such as NGC 5775, are now feasible and would provide much more information on the importance of secondary ionizing sources with local morphology and height.  It is important to emphasize that shocks are not the only possible secondary source.  Other sources such as TMLs, X-rays, and ionization by cosmic rays may contribute as well.

3. Another scenario that we consider is a pure photoionization model with an increasing temperature with $z$. By making a few reasonably valid assumptions, the run of gas temperature vs. $z$ has been determined in each of the four galaxies.  We find that an increase in temperature with $z$ can explain the general trend of an increase in the ratios of forbidden lines to Balmer lines with $z$, including an increase in [\ion{O}{3}]/H$\alpha$ and the constancy of [\ion{S}{2}]/[\ion{N}{2}].  By establishing the run of gas temperature, constraints on ionization fractions of Sulfur, Oxygen, and Hydrogen could then be established.  The [\ion{O}{3}]/H$\alpha$ ratios in NGC 5775 and UGC 10288 generally indicate an exceptionally high ionization fraction of O$^{++}$, that invalidates the assumption that N$^{+}$ and H$^{+}$ have equal ionization fractions, indicating that the line ratios cannot be explained by a temperature increase alone.    The runs of line ratios in NGC 891, where the ionization fraction of O$^{++}$ is still somewhat high, and NGC 4302 are better explained by an increase in temperature with $z$, though the absence of [\ion{O}{3}]$\lambda$5007 data for NGC 4302 makes a proper assessment difficult.

Finally, [\ion{O}{1}]$\lambda$6300, [\ion{O}{2}]$\lambda$3727, and [\ion{O}{3}]$\lambda$5007 data for a vertically oriented slit would provide additional information on ionization and excitation conditions in these DIG halos.

\acknowledgments
We thank the KPNO staff for their help in obtaining the data.  We also thank L. M. Haffner, R. Reynolds, and R. Benjamin for comments in the preparation of this paper.  This research has made use of the NASA/IPAC Extragalactic database (NED).  This work was partially supported by NSF grant AST-9986113.

\clearpage 
\begin{deluxetable}{lllcccc}
\tablecolumns{7}
\tablewidth{0pc}
\tablecaption{SUMMARY OF OBSERVATIONS \label{t1}}
\tablehead{
\colhead{Galaxy} & \multicolumn{2}{c}{Galaxy Center\tablenotemark{a}} & \colhead{Adopted} & \colhead{Offset along} & \colhead{Hours of} & \colhead{Noise\tablenotemark{c}} \\ 
\colhead{and Slit} & \colhead{R.A.} & \colhead{Dec.} & \colhead{Distance\tablenotemark{b}} & \colhead{Major axis} & \colhead{Integration} & \colhead{} \\
\colhead{} & \colhead{(J2000)} & \colhead{(J2000)} & \colhead{(Mpc)} & \colhead{} & \colhead{} & \colhead{}}
\startdata
NGC 5775 (Slit 1) & 14$^{h}$53$^{m}$57$\fs$6 & 03$\arcdeg$32$\arcmin$40$\arcsec$ & 24.8 & 32\arcsec\ NW & 6.0 & 2.2 \nl
NGC 5775 (Slit 2) & & & & 20\arcsec\ SE & 5.0 & 2.4 \nl
NGC 4302 & 12 \ 21 \ 42.5 & 14  36 05 & 16.8 & 24\arcsec\ S & 4.0 & 9.7  \nl
UGC 10288 (Slit 1) & 16 \ 14 \ 25.1 & -00  12 27 & 31.5 & 30\arcsec\ E & 6.17 & 2.5 \nl
UGC 10288 (Slit 2) & & & & 35\arcsec\ W & 5.33 & 3.6 \nl
NGC 891 & 02 \ 22 \ 33.1 & 42 20 48 & 9.6 & 100\arcsec\ N & 1.5 & 16.1 \nl 
\enddata
\tablenotetext{a}{The \ion{H}{1}\ kinematic center of NGC 5775 was determined by Irwin \markcite{}(1994).  Central positions of NGC 4302, UGC 10288, and NGC 891 are from de Vaucouleurs et al. \markcite{}(1991).}
\tablenotetext{b}{Adopted distance of NGC 5775 is from Irwin \markcite{}(1994).  Distances for NGC 4302, UGC 10288, and NGC 891 are from Tully \markcite{}(1988).}
\tablenotetext{c}{Before spatial averaging.  Noise units are 10$^{-19}$ erg cm$^{-2}$ s$^{-1}$ \AA$^{-1}$ arcsec$^{-2}$.}
\end{deluxetable}

\clearpage
\begin{deluxetable}{lcc}
\tablecolumns{3}
\tablewidth{0pc}
\tablecaption{DIG SCALE HEIGHTS \label{t2}}
\tablehead{\colhead{} & \colhead{Profile} & \colhead{$H_{em}$} \\ \colhead{Galaxy} & \colhead{Description} & \colhead{(pc)}}
\startdata
NGC 5775 & Slit 1 (NE) & 2300 \nl
NGC 5775 & Slit 1 (SW) & 2300 \nl
NGC 5775 & Slit 2 (NE) & 2500 \nl
NGC 5775 & Slit 2 (SW) & 1800 \nl
NGC 4302 & East & 700 \nl
NGC 4302 & West & 550 \nl
UGC 10288 & Slit 1 (S) & 320 \nl
UGC 10288 & Slit 2 (S) & 460 \nl
\enddata
\end{deluxetable}
     
\clearpage
\begin{deluxetable}{lcccc}
\tablecolumns{5}
\tablewidth{0pc}
\tablecaption{CONSTRAINTS ON IONIZATION FRACTIONS\tablenotemark{a} \label{t3}}
\tablehead{
\colhead{Galaxy} & \colhead{Gas Temperature} & \colhead{$\left(\frac{\mbox{S$^{+}$/S}}{\mbox{H$^{+}$/H}}\right)$} & \colhead{$\left(\frac{\mbox{O$^{++}$/O}}{\mbox{H$^{+}$/H}}\right)$} & \colhead{$\left(\frac{\mbox{H$^{+}$}}{\mbox{H}}\right)$}
 \\
\colhead{and Slit} & \colhead{(10$^{4}$ K)} & \colhead{} & \colhead{} & \colhead{}}
\startdata
NGC 5775 (Slit 1) & 0.64--0.95 & 0.30--0.60 & 0.15--0.70\tablenotemark{b} & n/a \nl
NGC 5775 (Slit 2) & 0.62--0.95 & 0.35--0.80 & 0.30--0.75\tablenotemark{b} & 0.65--0.85 \nl 
UGC 10288 (Slit 1) & 0.60--1.05 & 0.55--1.00 & 0.50--0.80\tablenotemark{b} & 0.45--0.65 \nl
UGC 10288 (Slit 2) & 0.60--0.95 & 0.50--1.00 & 0.50--0.80\tablenotemark{b} & 0.30--0.60 \nl
NGC 4302 & 0.66--1.00 & 0.35--0.60 & n/a & n/a \nl
NGC 891 & 0.65--1.05 & 0.40--0.65 & 0.13--0.30 & 0.55--0.80 \nl
\enddata
\tablenotetext{a}{Values of ionization fractions are valid only for the case of a pure photoionization model with an increasing temperature with $z$.  See text for other assumptions used to determine these values.}
\tablenotetext{b}{Nearly all data points fall within this range for NGC 5775 and UGC 10288. However, some measured values indicate an ionization fraction of O$^{++}$ relative to H$^{+}$ that is nearly 1.0.}
\end{deluxetable}

\end{document}